 \documentclass[final,3p,times]{elsarticle}

\usepackage{euscript,amsmath,amsfonts,graphicx,wrapfig,times,color,wrapfig,lipsum,amssymb,natbib}

 \usepackage[colorlinks=true]{hyperref}
\usepackage{cite}

 \usepackage{lineno}
 
 \newcommand{\D}{\displaystyle}
\newcommand{\e}{{\rm e}}

\renewcommand{\d}{{\rm d}}

\newcommand{\qq}{{\mathbf q}}

\newcommand{\ve}{\varepsilon}

\newcommand{\W}{{\mathbf W}}

\newcommand{\p}{{\mathbf p}}

\newcommand{\pd}{\partial}

\newcommand{\uu}{{\bf u}}

\newcommand{\U}{{\bf U}}
\newcommand{\pp}{{\bf p}}
\newcommand{\bomega}{{\boldsymbol \Omega}}

\newcommand{\bphi}{{\boldsymbol \Phi}}

\newcommand{\mc}{\mathcal }

\journal{Physica D}

\begin{document}

\begin{frontmatter}



\title{Delay stabilizes stochastic motion of bumps in layered neural fields}


\author{Zachary P. Kilpatrick}
\ead{zpkilpat@math.uh.edu}
\address{Department of Mathematics, University of Houston, Houston, TX 77204}

\begin{abstract} We study the effects of propagation delays on the stochastic dynamics of bumps in neural fields with multiple layers. In the absence of noise, each layer supports a stationary bump. Using linear stability analysis, we show that delayed coupling between layers causes translating perturbations of the bumps to decay in the noise-free system. Adding noise to the system causes bumps to wander as a random walk. However, coupling between layers can reduce the variability of this stochastic motion by canceling noise that perturbs bumps in opposite directions. Delays in interlaminar coupling can further reduce variability, since they couple bump positions to states from the past. We demonstrate these relationships by deriving an asymptotic approximation for the effective motion of bumps. This yields a stochastic delay-differential equation where each delayed term arises from an interlaminar coupling. The impact of delays is well approximated by using a small delay expansion, which allows us to compute the effective diffusion in bumps' positions, accurately matching results from numerical simulations. \end{abstract}

\begin{keyword} neural field equations \sep delay differential equations \sep effective diffusion




\end{keyword}

\end{frontmatter}

\section{Introduction}
\label{intro}


Delays commonly arise in dynamical models of large scale neuronal networks, often accounting for the detailed kinetics of chemical or electrical activity \citep{stepan09}. The finite-velocity of action potential (AP) propagation can lead to delays on the order of milliseconds between AP instantiation at the axon hillock and its arrival at the synaptic bouton \citep{stuart97}.  Similar propagation delays have been observed in dendritic APs propagating to the soma \citep{vetter01}. Furthermore, synaptic processing involves several steps including vesicle release, neurotransmitter diffusion, and uptake, so the chemical signal communicating between cells is effectively delayed \citep{markram97}. However, computational models of large scale networks that describe all these processes in detail are unwieldy, not admitting direct analysis, so one must rely on expensive simulations to study their behavior \citep{izhikevich08}. An alternative approach is to develop mean field models of spiking networks that incorporate delay that accounts for these microscopic processes \citep{roxin05}.

Neural field equations are a canonical model of large scale spatiotemporal activity in the brain \citep{bressloff12}. Many studies have explored the impact of delays on the resulting spatiotemporal solutions of these equations \citep{pinto01,coombes03,hutt03}. One common observation is that the inclusion of delays can lead to oscillations via a Hopf bifurcation in the linear system describing the local stability of solutions to the delay-free system: Turing patterns \citep{hutt03}, stationary pulses \citep{veltz13,faye14}, and traveling waves \citep{roxin05,laing06}. Thus, a major finding across many studies of delayed neural field equations is that delay will tend to contribute to instabilities in stationary states \citep{coombes09}. Recent work has shown that in stochastic neural field models, delay can stabilize the system near bifurcations \citep{hutt12}. This distinction has been explored extensively in control theory literature: delayed negative feedback loops can induce instability while delayed positive feedback can augment stability \citep{abdallah93}. In this work, we further explore the potential stabilizing impact of delays in neural field models. Specifically, we focus on the case where positive feedback between two layers of a neural field help stabilize patterns to noise perturbations.

We will focus specifically on a multilayer neural field model that supports {\em bump attractors} \citep{kilpatrick13c}. Persistent spiking activity with a ``bump" shape is an experimentally observed neural substrate of spatial working memory \citep{funahashi89,wimmer14}. The position of the bump encodes the remembered location of a cue \citep{compte00}. Noise degrades memory accuracy over time \citep{laing01}, due to diffusive wandering of bumps across the neutrally stable landscape of the network \citep{amari77}. Several mechanisms have been proposed to limit such diffusion-induced error: short term facilitation \citep{itskov11,hansel13}, bistable neural units \citep{camperi98,koulakov02}, and spatially heterogeneous recurrent excitation \citep{kilpatrick13,kilpatrick13b}. Recently, we showed interlaminar coupling, known to exist between the many brain areas participating in spatial working memory \citep{curtis06}, can also help to reduce bump position variability due to noise cancellation. Here, we show that delays in the interlaminar coupling further reduce the long term variability in bump positions. Essentially, this occurs because each layer is constantly coupled to past states of other layers, states that have integrated noise for a shorter length of time than the current state.

The paper is organized as follows. In section \ref{model}, we introduce the multilayer neural field model with delays and noise, showing they take the form of a delayed stochastic integrodiffferential equation. Section \ref{bumpmot} then explores how delays impact the local stability of stationary bumps in a dual layer neural field, in the absence of noise. Essentially, we demonstrate the delay reduces the impact of translating perturbations to the bump solution, underlying the mechanism of position stabilization. This motivates our findings in section \ref{stochmot}, where we derive effective stochastic equations for the motion of bump solutions subject to noise, showing they take the form of stochastic delay differential equations. A small delay expansion allows us to compute an effective variance, which is shown to be reduced by increasing the delay in coupling between layers. Lastly, we extend our results in section \ref{multi}, showing similar results hold in stochastic neural fields with more than two layers, and the effective variance decreases with the number of layers.

\section{Laminar neural fields with delays and noise}
\label{model}
\subsection{Dual layer neural field with delays between layers}

We model a pair of reciprocally coupled stochastic neural fields, accounting for the the propagation delay between layers as:
\begin{subequations} \label{delayers}
\begin{align}
\d u_1(x,t) &= \left[ - u_1(x,t) + \int_{- \pi}^{\pi} w(x-y) f(u_1(y,t)) \d y + \int_{- \pi}^{\pi} w_{12} (x-y) f(u_2(y,t - \tau_{12} (x,y))) \d y \right] \d t + \ve \d W_1(x,t), \\
\d u_2 (x,t) &= \left[ - u_2 (x,t) + \int_{- \pi}^{\pi} w(x-y) f(u_2(y,t)) \d y + \int_{- \pi}^{\pi} w_{21}(x-y) f(u_1(y,t - \tau_{21}(x,y))) \d y \right] \d t + \ve \d W_2(x,t),
\end{align}
\end{subequations}
so $u_j(x,t)$ is the total synaptic input at location $x \in [ - \pi, \pi]$ in layer $j$. The effects of synaptic architecture are given by the convolution terms, so $w(x-y)$ describes the polarity (sign of $w$) and strength (amplitude of $w$) of recurrent connectivity within a layer. Typically, bump attractor network models assume spatially dependent synaptic connectivity that is lateral inhibitory \citep{amari77}, such as the cosine
\begin{align}
w(x-y) = \cos (x-y), \hspace{5mm} j=1,2,  \label{wcos}
\end{align}
but our analysis will apply to the general case of any even weight function. Synaptic connections from layer $k$ to $j$ are described by the kernels $w_{jk}(x-y)$. To compare our analysis with numerical simulations, we will use the cosine coupling
\begin{align}
w_{jk}(x-y) = M_j \cos (x-y), \hspace{5mm} k \neq j,  \label{intercos}
\end{align}
where $M_j$ specifies the strength of coupling projecting to the $j$th layer.

Another feature of long range coupling is that the activity signals can take a finite amount of time to propagate from one neuron to the next \citep{manor91,vetter01,debanne04}. Thus, delay is incorporated into the connectivity between layers through the spatially dependent functions $\tau_{jk}(x,y)$ \citep{bressloff96,hutt03,coombes03,roxin05}, describing the amount of time it takes a signal to propagate from location $y$ in layer $k$ to location $x$ in layer $j$. Our analysis can be carried out in the case of general functions $\tau_{jk}(x,y)$, but we demonstrate our results using specific cases, such as hard delays $\tau_{jk}(x,y) = \bar{\tau}_{jk}$ (constant) or distance-dependent delays (e.g., $\tau_{jk}(x,y) = \tilde{\tau}_{jk}(x-y)$). 

Firing rate functions $f(u)$ are typically nonlinear monotonic functions of the synaptic input $u$, which we take to be sigmoidal \citep{wilson73}
\begin{align*}
f(u) = \frac{1}{1 + \e^{- \gamma ( u - \theta )}},
\end{align*}
with threshold $\theta$ and gain $\gamma$. To compute quantities explicitly, we typically take the high gain limit $\gamma \to \infty$ to yield the Heaviside firing rate function \citep{amari77}
\begin{align}
f(u) = H(u - \theta) = \left\{ \begin{array}{ll} 0 & : u < \theta, \\ 1 & : u \geq \theta.   \end{array} \right.  \label{H}
\end{align}
Noise in each layer $j$ is described by a small amplitude ($0 \leq \ve \ll 1$) stochastic process $\d W_j (x,t)$ that is white in time and correlated in space so that $\langle \d W_j (x,t) \rangle = 0$ and
\begin{align*}
\langle \d W_j(x,t) \d W_j (y,s) \rangle &= C_j(x-y) \delta ( t- s) \d t \d s, \\
\langle \d W_j(x,t) \d W_k (y,s) \rangle &= C_c (x-y) \delta (t-s) \d t \d s,
\end{align*}
describing both local ($C_j(x-y)$, $j=1,2$) and shared ($C_c(x-y)$) noise correlations as a function of the difference in positions. Notice, in the case $C_c \equiv 0$, there are no inter laminar noise correlations, whereas if $C_1 \equiv C_2 \equiv C_c$, noise in each layer is drawn from the same process. In explicit examples, we typically take cosine spatial correlation functions
\begin{align}
C_j(x) = c_j \cos (x), \hspace{5mm} C_c = c_c \cos (x).  \label{coscorr}
\end{align}

\subsection{Multiple layer neural field with delays between layers}

We can extend our neural field model with two delay-coupled layers to an arbitrary number of layers $N$ with any synaptic architecture in between, as described by the system of stochastic integrodifferential equations
\begin{align}
\d u_j(x,t) &= \left[ - u_j(x,t) + \int_{- \pi}^{\pi} w(x-y) f(u_j(y,t)) \d y + \sum_{k \neq j} \int_{- \pi}^{\pi} w_{jk} (x-y) f(u_k(y,t - \tau_{jk} (x,y))) \d y \right] \d t + \ve \d W_j(x,t), \label{multilayers}
\end{align}
where $u_j(x,t)$ is neural activity in the $j$th layer ($\forall j=1,...,N$), and the notation $\sum_{k \neq j} \equiv \sum_{k=1, k \neq j}^N$. Connectivity between layers is described by the synaptic weight function $w_{jk}(x-y)$ linking position $y$ in layer $k$ to position $x$ in layer $j$. For comparison with numerical simulations, we will utilize cosine shaped connectivity (\ref{wcos},\ref{intercos}) and a Heaviside firing rate function (\ref{H}). As in our model with two layers, noises $W_j(x,t)$ are white in time and correlated in space so $\langle \d W_j(x,t) \rangle = 0$ and
\begin{align*}
\langle \d W_j(x,t) \d W_k(y,t) \rangle = C_{jk}(x-y) \delta (t-s) \d t \d s,
\end{align*}
with $\forall j,k=1,...,N$. For comparison with numerics, local ($j=k$) correlations will use the correlation function $C_{jj}(x) = \cos (x)$ and interlaminar ($j \neq k$) correlations will take $C_{jk}(x) = c_c \cos (x)$ for all $j \neq k$.

\section{Impact of delays on bump stability}
\label{bumpmot}

\begin{figure}[tb]
\begin{center}  \includegraphics[width=7.6cm]{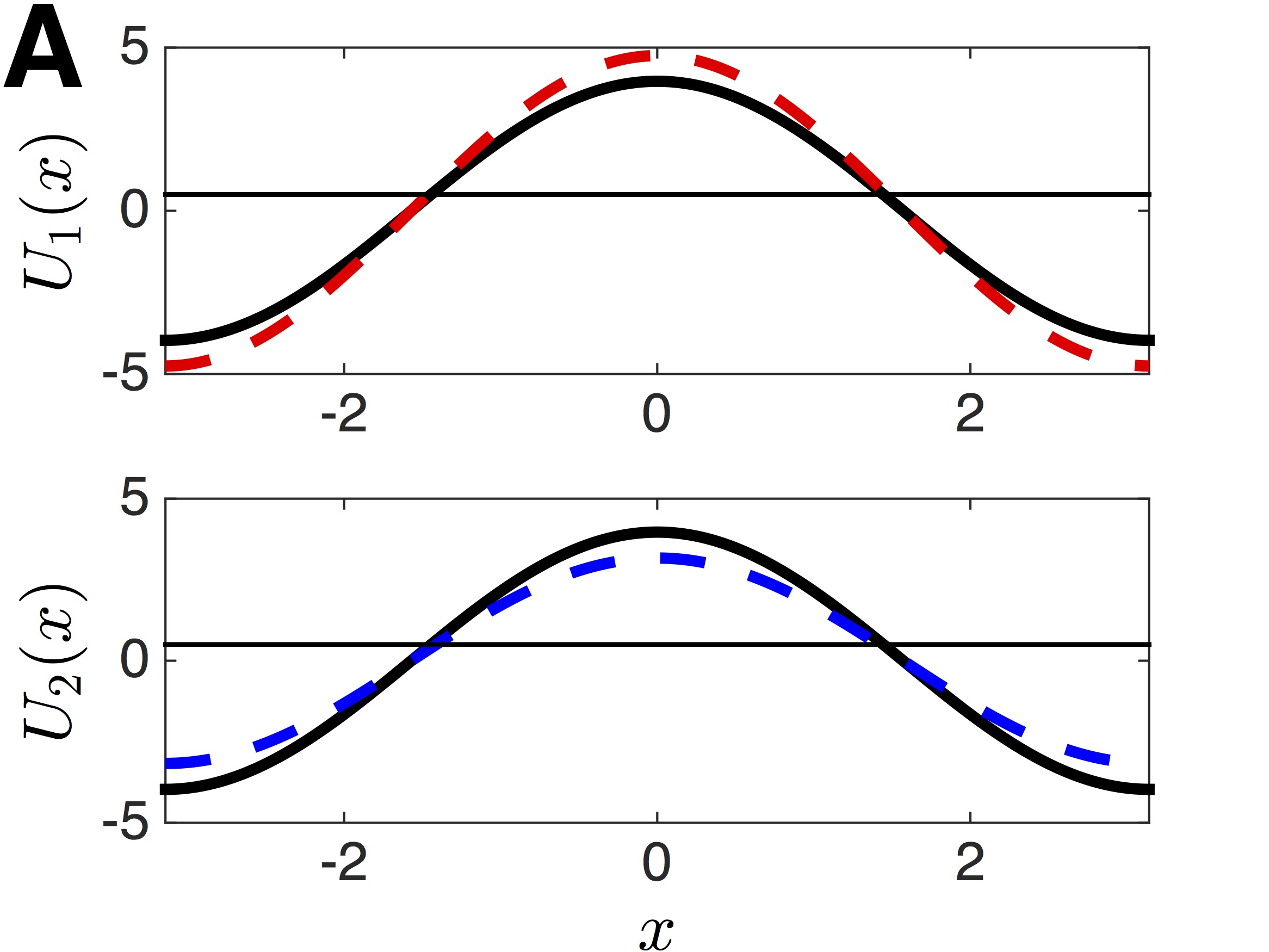}  \includegraphics[width=8.1cm]{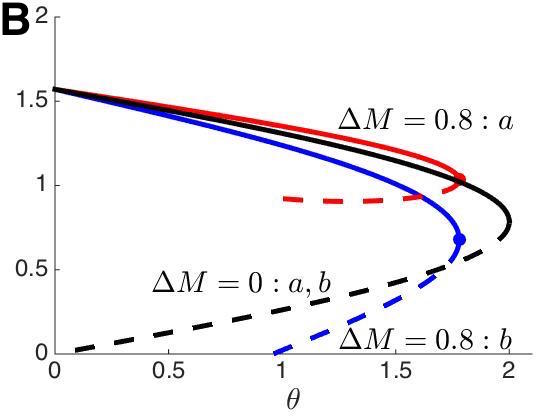} \end{center}
\caption{({\bf A}) Profiles of the coupled stable bump solutions $(U_1(x),U_2(x))$ are identical (solid curves) when coupling strength is symmetric ($w_{12}(x) = w_{21}(x) \equiv \cos (x)$). However, when layer 1 receives stronger coupling than layer 2 ($w_{12}(x) = 1.4 \cos (x) , w_{21}(x) = 0.6 \cos (x)$), the bump in layer 1 ($U_1(x)$) is larger than that in layer 2 ($U_2(x)$) (dashed lines).  Threshold (thin line) $\theta = 0.5$. ({\bf B}) As the threshold $\theta$ is increased, the wide (solid) and narrow (dashed) solution branches vary until coalescing in a saddle-node bifurcation (filled dot). Half-widths $a$ and $b$ are identical when coupling is symmetric ($M_1 = M_2 \equiv 1$ and $\Delta M = M_1 - M_2 = 0$). Notice the stable and wide branch of solutions increases width when layer 1 receives more input ($M_1 = 1.4$ and $M_2 = 0.6$). Local connectivity $w(x) = \cos (x)$.}
\label{cupbwid}
\end{figure}

We are interested in how delays and coupling impact the stability of the stationary bump solutions $(U_1(x),U_2(x))$, as this will foreshadow how noise will impact their perturbative motion. Rather than carrying out an exhaustive study of the spectrum of the linearized operator about the bump solution, we will focus on how delays impact the stability of the bump to translating perturbations. Bumps are well accepted models of persistent working memory, so their position represents a memory of their initial condition \citep{goldmanrakic95,compte00,kilpatrick13b}.  Our main goal will be to demonstrate that bump positions are displaced a shorter distance in neural field layers with reciprocal delayed coupling. It is well known that bump solutions in translationally symmetric neural fields have positions that lie upon a line attractor, so they are neutrally stable to perturbations that change their position \citep{amari77,camperi98,ermentrout98}. Previously we showed that weak interlaminar coupling decreases the overall displacement of bumps by spatiotemporal noise, since perturbations that move bumps in the opposite direction are canceled \citep{kilpatrick13c}.

We begin by explicitly calculating bump solutions to the dual layer model (\ref{delayers}) with arbitrarily strong coupling. Note, a similar study was carried out recently, in the absence of noise on an infinite domain \citep{folias11}. We begin by considering the noise-free case, so $\d W_j \equiv 0$, $j=1,2$. We can thus determine the form of coupled stationary bump solutions $(u_1,u_2) = (U_1(x),U_2(x))$ self consistently, so they satisfy the stationary equation
\begin{align}
U_1(x) &= \int_{- \pi}^{\pi} w (x-y) f(U_1(y)) \d y + \int_{- \pi}^{\pi} w_{12}(x-y) f(U_2(y)) \d y, \nonumber \\
U_2 (x) &= \int_{- \pi}^{\pi} w(x-y) f(U_2(y)) \d y + \int_{- \pi}^{\pi} w_{21}(x-y) f(U_1(y)) \d y.  \label{bdualint}
\end{align}
Notice that the delays do not impact the form of the bump solution, since they are determined by a stationary equation. Assuming even symmetric weight functions and a Heaviside firing rate function (\ref{H}) allows us to fix the threshold crossing points of bumps, so that $U_1( \pm a ) = \theta$ and $U_2 (\pm b) = \theta$. In the parlance of \citep{folias11}, we shall only examine {\em syntopic} bumps (bump centered at the same location in each layer). This converts the implicit integral equation system (\ref{bdualint}) to an explicit expression for both bumps
\begin{align}
U_1(x) = \int_{-a}^{a} w(x-y) \d y + \int_{- b}^{b} w_{12} (x-y) \d y,  \hspace{9mm} U_2(x) = \int_{-b}^{b} w(x-y) \d y + \int_{-a}^{a} w_{21} (x-y) \d y,  \label{bsols}
\end{align}
where we now need only determine the bump half-widths $a$ and $b$. We can do so, by requiring self-consistency of the expressions $U_1(a) = \theta$ and $U_2(b) = \theta$, so
\begin{align}
\theta = \int_0^{2a} w(x) \d x + \int_{a-b}^{a+b} w_{12} (x) \d x, \hspace{9mm} \theta = \int_0^{2b} w(x) \d x + \int_{b-a}^{a+b} w_{21} (x) \d x. \label{genabeqn}
\end{align}
Upon considering cosine weight functions (\ref{wcos},\ref{intercos}), we find (\ref{genabeqn}) integrates to
\begin{align*}
\theta = 2 \cos(a) [ \sin(a) + M_1 \sin (b)], \hspace{9mm} \theta = 2 \cos(b) [\sin(b) + M_2 \sin (a)].
\end{align*}
We demonstrate the relationship between the bump half-widths $a$ and $b$ and the threshold $\theta$ as well as the coupling amplitudes $M_1$ and $M_2$ in Fig. \ref{cupbwid}. Note that in the symmetric case $M_1 = M_2 \equiv M$, we have $a=b$, so
\begin{align*}
\theta = 2 (1+M) \cos (a) \sin (a),
\end{align*}
which can be solved to yield two solutions, a wide ($a_w$) and narrow ($a_n$) bump pair
\begin{align*}
a_w = \frac{\pi}{2} - \frac{1}{2} \sin^{-1} \frac{\theta}{1+M}, \hspace{8mm} a_n = \frac{1}{2} \sin^{-1} \frac{\theta}{1+M}.
\end{align*}
These two solution branches will annihilate one another when $\theta = 1+M$. Thus, notice that interlaminar coupling expands the region of parameter space in which bumps exist.

Now, we analyze linear stability by studying the evolution of small, smooth, and separable perturbations to the bumps given by the functions $\ve \psi_j(x,t)$ ($\ve \ll 1$), $j=1,2$. We derive this linearization by employing the expansion
\begin{align}
u_1(x,t) &= U_1(x) + \ve \psi_1 (x,t) + {\mc O}(\ve^2), \nonumber \\
u_2(x,t) &= U_2(x) + \ve \psi_2 (x,t) + {\mc O}(\ve^2). \label{stabexpan}
\end{align}
Plugging this expansion into (\ref{delayers}), in the absence of noise ($\d W_j \equiv 0$, $j=1,2$), and truncating to ${\mc O}(\ve)$, we find $(\psi_1(x,t), \psi_2(x,t))$ satisfy the system
\begin{align}
\dot{\psi}_1 (x,t) &= - \psi_1(x,t) + \int_{- \pi}^{\pi} w(x-y) f'(U_1(y)) \psi_1(y,t) \d y + \int_{- \pi}^{\pi} w_{12}(x-y) f'(U_2(y)) \psi_2(y,t - \tau_{12}(x,y)) \d y, \nonumber \\
\dot{\psi}_2(x,t) &= - \psi_2(x,t) + \int_{- \pi}^{\pi} w(x-y) f'(U_2(y)) \psi_2(y,t) \d y + \int_{- \pi}^{\pi} w_{21} (x-y) f'(U_1(y)) \psi_1(y,t - \tau_{21}(x,y)) \d y,  \label{psifulsys}
\end{align}
where $\dot{\psi}_j = \pd_t \psi_j(x,t)$ ($j=1,2$). We can immediately identify the neutrally stable solution given by the derivative $(\psi_1(x,t), \psi_2(x,t)) = (U_1'(x),U_2'(x))$ by simply plugging this ansatz into (\ref{psifulsys}) to yield
\begin{align}
0 &= - U_1'(x) + \int_{- \pi}^{\pi} w(x-y) f'(U_1(y)) U_1'(y) \d y + \int_{- \pi}^{\pi} w_{12}(x-y) f'(U_2(y)) U_2'(y) \d y, \nonumber \\
0 &= - U_2'(x) + \int_{- \pi}^{\pi} w(x-y) f'(U_2(y)) U_2'(y) \d y + \int_{- \pi}^{\pi} w_{21}(x-y) f'(U_1(y)) U_1'(y) \d y. \label{Udeqns}
\end{align}
The fact that (\ref{Udeqns}) holds can be seen by differentiating the system (\ref{bdualint}) and using integration by parts to rearrange the integral terms. Similar results have been founded in linear stability analyses of non-delayed neural field equations, and they typically imply that perturbations that translate solutions in precisely this way will neither grow nor decay \citep{bressloff01,coombes04,kilpatrick13}. However, we will demonstrate that this result is misleading in the delayed case. In fact, instantaneous perturbations of this form may decay, and the stabilizing impact of propagation delays relies on this subtle difference.

To analyze the dynamics of (\ref{psifulsys}) in more detail, we first simplify the system, assuming a Heaviside firing rate function (\ref{H}). This allows us to examine the dynamics of the perturbations $\psi_1$ and $\psi_2$ at single points $x= \pm a$ and $x= \pm b$ respectively.  In this case, we can compute
\begin{align*}
f'(U_1) = \gamma_a [\delta (x-a) + \delta (x+a)], \hspace{9mm} f'(U_2) = \gamma_b [\delta (x-b) + \delta (x+b)],
\end{align*}
where
\begin{align}
\gamma_a^{-1} &= |U_1'(a)| = |U'(-a)| = w(0) - w(2a) + w_{12}(b-a) - w_{12}(a+b), \nonumber \\
\gamma_b^{-1} &= |U_2'(b)| = |U'(-b)| = w(0) - w(2b) + w_{21}(b-a) - w_{21}(a+b). \label{gamab}
\end{align}
The integrals in (\ref{psifulsys}) can then be calculated so that
\begin{align}
\dot{\psi}_1 (x,t) &= - \psi_1(x,t) + \gamma_a \sum_{\D x_a = \pm a} w(x-x_a) \psi_1(x_a,t) + \gamma_b \sum_{\D x_b = \pm b} w_{12}(x-x_b) \psi_2(x_b,t - \tau_{12}(x,x_b)), \nonumber \\
\dot{\psi}_2 (x,t) &= - \psi_2(x,t) + \gamma_b \sum_{\D x_b = \pm b} w(x-x_b) \psi_2(x_b,t) + \gamma_a \sum_{\D x_a = \pm a} w_{12}(x-x_a) \psi_1(x_a,t - \tau_{21}(x,x_a)). \label{psidsums}
\end{align}
The essential spectrum of the linearized system (\ref{psidsums}) is associated with solutions of the form $\psi_1(\pm a,t) = \psi_2( \pm b, t) \equiv 0$ ($\forall t$) and $\psi_j(x,t) = \e^{- t} \bar{\psi}(x)$, which does not contribute to any instabilities. Perturbations of other forms can be studied by focusing on the values $\psi_1(\pm a, t)$ and $\psi_2( \pm b, t)$, which satisfy the delayed system of differential equations
\begin{align}
\dot{\psi}_1 (-a,t) &= - \psi_1(-a,t) + \gamma_a \sum_{\D x_a = \pm a} w(-a-x_a) \psi_1(x_a,t) + \gamma_b \sum_{\D x_b = \pm b} w_{12}(-a-x_b) \psi_2(x_b,t - \tau_{12}(-a,x_b)), \nonumber \\
\dot{\psi}_1 (a,t) &= - \psi_1(a,t) + \gamma_a \sum_{\D x_a = \pm a} w(a-x_a) \psi_1(x_a,t) + \gamma_b \sum_{\D x_b = \pm b} w_{12}(a-x_b) \psi_2(x_b,t - \tau_{12}(a,x_b)), \nonumber \\
\dot{\psi}_2 (-b,t) &= - \psi_2(-b,t) + \gamma_b \sum_{\D x_b = \pm b} w(-b-x_b) \psi_2(x_b,t) + \gamma_a \sum_{\D x_a = \pm a} w_{21}(-b-x_a) \psi_1(x_a,t - \tau_{21}(-b,x_a)), \nonumber \\
\dot{\psi}_2 (b,t) &= - \psi_2(b,t) + \gamma_b \sum_{\D x_b = \pm b} w(b-x_b) \psi_2(x_b,t) + \gamma_a \sum_{\D x_a = \pm a} w_{21}(b-x_a) \psi_1(x_a,t - \tau_{21}(b,x_a)). \label{psidsums}
\end{align}
Furthermore, we can specifically examine how the width and position of bumps changes by studying the four threshold crossing points satisfying
\begin{align}
u_1(\pm a + \ve \alpha_{\pm}(t),t) = \theta + {\mc O}(\ve^2),  \hspace{9mm} u_2( \pm b + \ve \beta_{\pm}(t),t) = \theta + {\mc O}(\ve^2), \label{tpabt}
\end{align}
since perturbations are ${\mc O}(\ve)$. Thus, by Taylor expanding (\ref{tpabt}) and applying the ansatz (\ref{stabexpan}), we find at ${\mc O}(\ve)$
\begin{align}
\alpha_{\pm} (t) = \pm \gamma_a \psi_1(\pm a,t), \hspace{9mm} \beta_{\pm}(t) = \pm \gamma_b \psi_2(\pm b,t).  \label{alfbet}
\end{align}
Substituting the expressions (\ref{alfbet}) into the system (\ref{psidsums}) and considering the case where $\tau_{12}$ and $\tau_{21}$ are distance-dependent so $\tau_{12}(x,y) = \tilde{\tau}_{12}(|x-y|)$ and $\tau_{21}(x,y) = \tilde{\tau}_{21}(|x-y|)$ \citep{hutt03,coombes03}, we find
\begin{align}
\dot{\alpha}_-(t) &= - \alpha_-(t) + \gamma_a \left[ w(0) \alpha_-(t) - w(2a) \alpha_+(t)  + w_{12}(b-a) \beta_- (t - \tilde{\tau}_{12}(|b-a|)) -  w_{12}(a+b) \beta_+(t - \tilde{\tau}_{12}(a+b)) \right], \nonumber \\
\dot{\alpha}_+(t) &= - \alpha_+ (t) + \gamma_a \left[ -w(2a) \alpha_-(t) + w(0) \alpha_+(t) - w_{12}(a+b) \beta_-(t - \tilde{\tau}_{12}(a+b)) + w_{12}(b-a) \beta_+(t - \tilde{\tau}_{12}(|b-a|)) \right], \nonumber \\
\dot{\beta}_-(t) &= - \beta_-(t) + \gamma_b \left[ w(0) \beta_-(t) - w(2b) \beta_+(t) + w_{21}(b-a) \alpha_-(t - \tilde{\tau}_{21}(|b-a|)) - w_{21}(a+b) \alpha_+(t - \tilde{\tau}_{21}(a+b)) \right], \nonumber \\
\dot{\beta}_+(t) &= - \beta_+(t) + \gamma_b \left[ -w(2b) \beta_-(t) + w(0) \beta_+(t) - w_{21}(a+b) \alpha_-(t - \tilde{\tau}_{21}(a+b)) + w_{21}(b-a) \alpha_+(t - \tilde{\tau}_{21}(|b-a|)) \right]. \label{alfbetsys}
\end{align}
Our main concern is the impact of delays on the stability of the bump solution's position. Assuming the long term width of the bump stays the same ($\lim_{t \to \infty} \alpha_+(t) = \lim_{t \to \infty} \alpha_- (t)$ and $\lim_{t \to \infty} \beta_+(t) = \lim_{t \to \infty} \beta_- (t)$), we can determine the long term position of the bump by studying the evolution of the summed variables $\alpha (t) := (\alpha_+(t) + \alpha_-(t))/2$ and $\beta (t) := (\beta_+(t) + \beta_-(t))/2$. By summing equations of the system (\ref{alfbetsys}), we find that
\begin{align}
\dot{\alpha}(t) &= - (W_{-1} +W_{+1}) \alpha (t) + W_{-1} \beta (t - T_{-1}) + W_{+1} \beta ( t - T_{+1}), \nonumber \\
\dot{\beta}(t) &= - (W_{-2} + W_{+2}) \beta (t) + W_{-2} \alpha (t - T_{-2}) + W_{+2} \alpha ( t - T_{+2}),  \label{abdelsys}
\end{align}
where $W_{\pm 1} := \gamma_a w_{12}(b \pm a)$, $W_{\pm 2} := \gamma_b w_{21}(b \pm a)$, $T_{\pm 1} := \tilde{\tau}_{12}(|b \pm a|)$, and $T_{\pm 2} := \tilde{\tau}_{21}(|b \pm a|)$. Instantaneous perturbations of the positions $\alpha (t)$ and $\beta (t)$ will always decay slightly in the limit when the effective delays are positive ($T_{\pm 1}, T_{\pm 2} > 0$). That is $\lim_{t \to \infty} \alpha (t) < \alpha (0)$ and $\lim_{t \to \infty} \beta (t) < \beta (0)$. 

\begin{figure}[tb]
\begin{center}  \includegraphics[width=8cm]{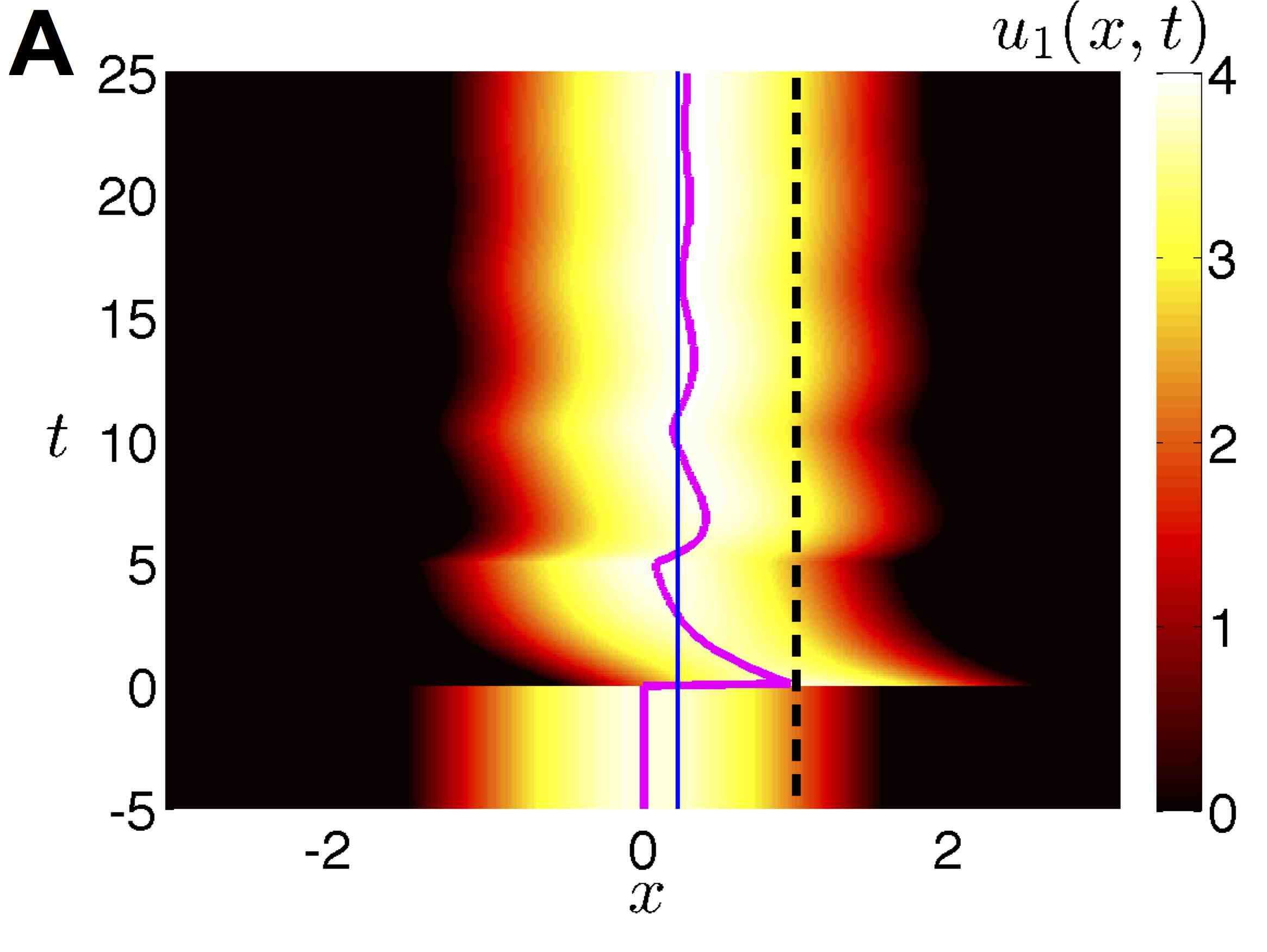} \includegraphics[width=8cm]{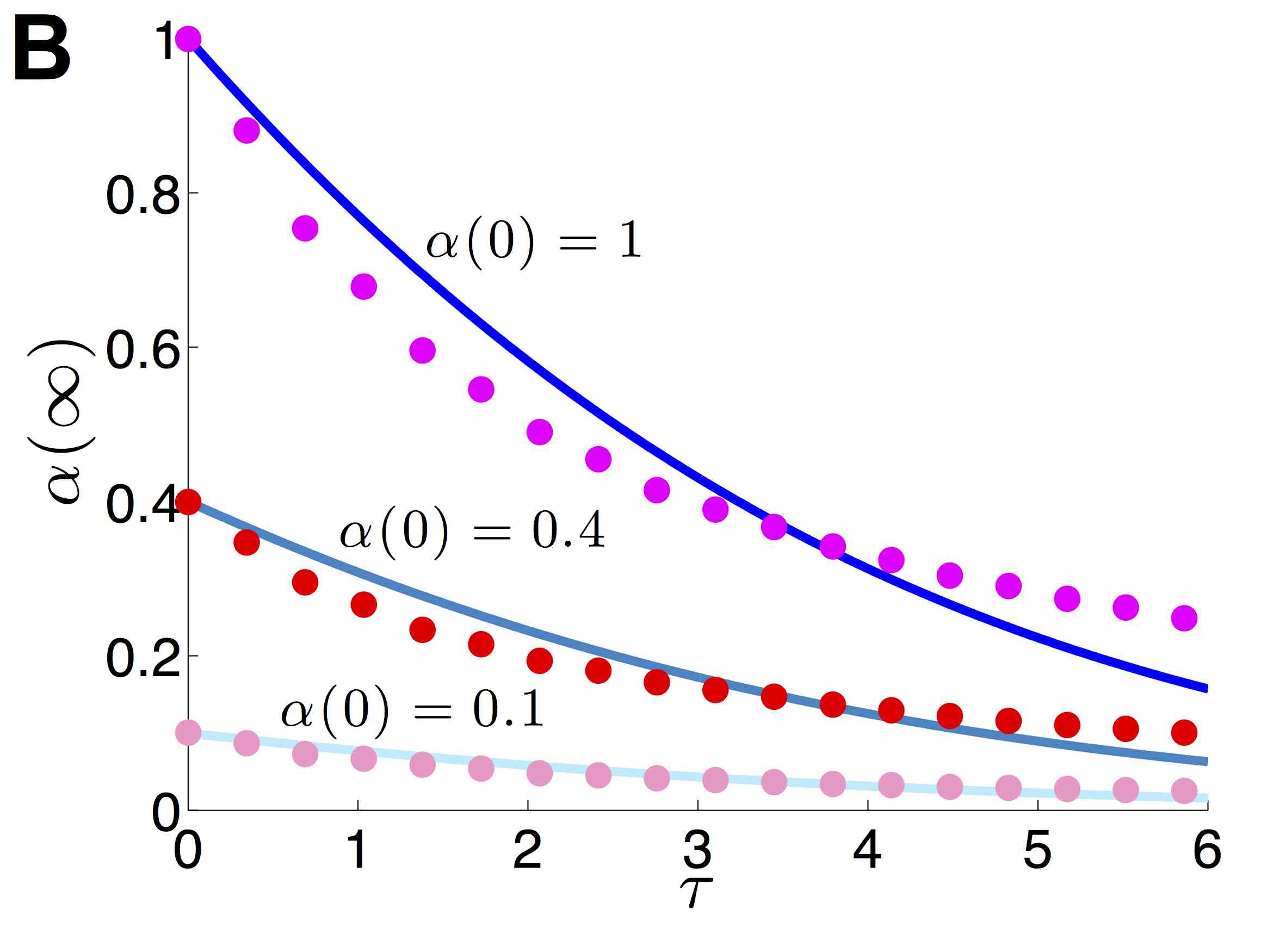} \end{center}
\caption{({\bf A}) Response of bump solution (\ref{bsols}) of the system (\ref{delayers}) to an instantaneous shift perturbation (dashed line) $\alpha_{\pm}(0)=\beta_{\pm}(0)= 1$ (defined in (\ref{tpabt})). Delayed coupling between layers ($\tau_{12}=\tau_{21}\equiv \tau = 5$) reduces the long term impact of the perturbations (thick line) as predicted by theory (thin line) in (\ref{delscale}). ({\bf B}) Theoretical predictions (solid lines) of the long term shift $\alpha ( \infty ) = \beta ( \infty )$ of the bump solution due to an initial shift with magnitude $\alpha (0) = \beta (0)$ matches numerical simulations (dots) of (\ref{delayers}). As the delay $\tau_{12}(x,y) = \tau_{21}(x,y) \equiv \tau$ in coupling between layers increases, the long term shift is reduced. Threshold $\theta = 0.5$ and couplings $w_{12}(x) = w_{21}(x) = \cos (x)$.}
\label{delpert}
\end{figure}

We demonstrate the precise amount by which delays reduce translating perturbations of bump position in the straightforward case of symmetric coupling ($W_{\pm 1} \equiv W_{\pm 2}$) and symmetric and hard delays ($T_{\pm1} \equiv T_{\pm 2} \equiv T$). In this case, the four lag system (\ref{abdelsys}) becomes a symmetric single lag system
\begin{align}
\dot{\alpha}(t) = W_T \left[ \beta ( t - T)  - \alpha (t) \right], \hspace{9mm} \dot{\beta}(t) = W_T \left[ \alpha ( t - T) - \beta (t) \right],
\end{align}
where $W_T := W_{- 1} + W_{+1} = W_{-2} + W_{+2}$. For initial conditions $\alpha (0) = \beta (0)$, and $\alpha (t) =\beta (t) =0 $ for $t \in ( - \infty, 0)$, it is straightforward to calculate that $\alpha (t) = \beta (t) = \alpha (0) \e^{-W_T t}$ on $t \in [0,T]$. Subsequently, we can solve $\dot{\alpha} = - W_T \alpha (t) + \alpha (0) W_T \e^{- W_T t}$ on $t \in [T,2T]$ to yield $\alpha (t) = \alpha (0) \e^{-2 W_T (t- T)} ( \e^{-W_T T} + W_T ( t - T)) $ for $t \in [T, 2T]$, as well as an identical result for $\beta$. Iterating this process, we find
\begin{align}
\lim_{t \to \infty} \alpha (t) = \lim_{t \to \infty} \beta (t) = \alpha (0) W_T T \sum_{n=1}^{\infty} \e^{-n W_T T} = \frac{\alpha (0) W_T T}{\e^{W_T T}  - 1} < \alpha (0),  \label{delscale}
\end{align} 
for $W_T T > 0$, so delay reduces the distance the bump will be perturbed compared to the case of no coupling or delay
\begin{align*}
\lim_{W_T T \to 0} \frac{\alpha (0) W_T T}{\e^{W_T T}  - 1} = \alpha (0).
\end{align*}
We demonstrate this effect in a simulation as well as with plots of the theoretical prediction (\ref{delscale}) in Fig. \ref{delpert}.

\section{Stochastic motion of bumps in dual layer network with delays}
\label{stochmot}
 
\subsection{Effective equations for stochastic bump motion}
\label{effeqns}

\begin{figure}[tb]
\begin{center} \includegraphics[width=9.4cm]{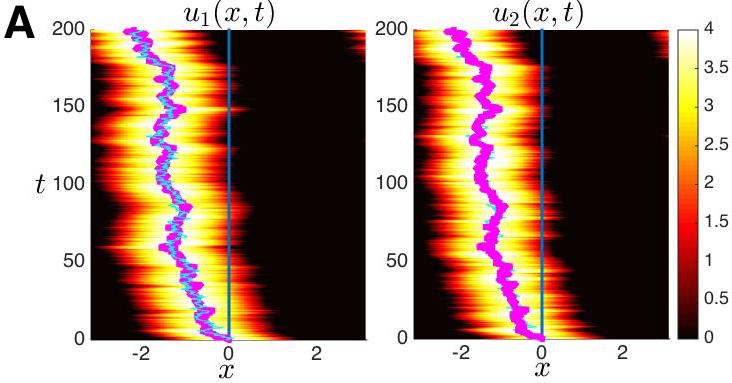} \includegraphics[width=6.6cm]{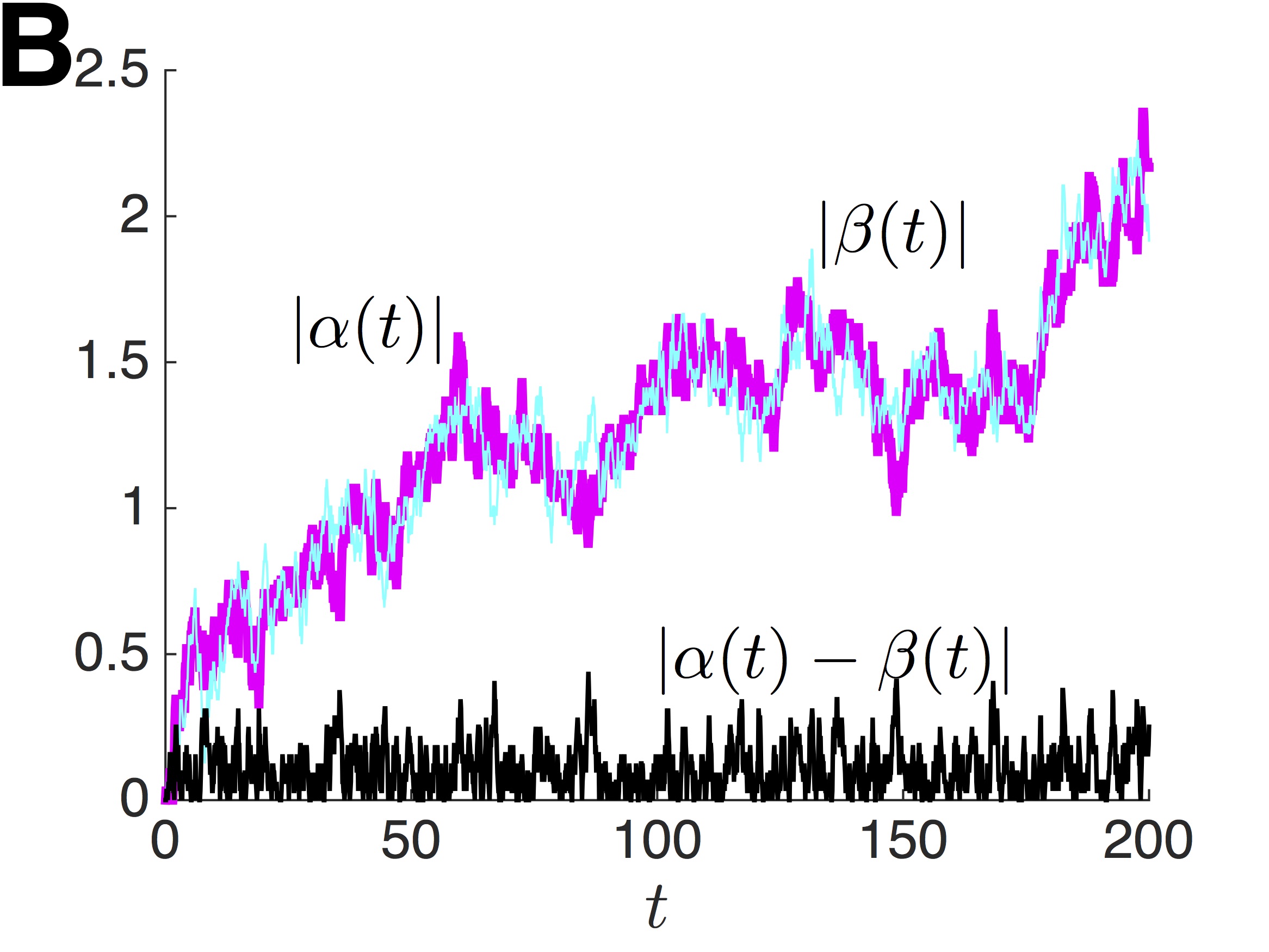} \end{center}
\caption{({\bf A}) Realization of the stochastic neural field (\ref{delayers}) reveals how delayed coupling between layers tends to keep positions of bumps in layers 1 and 2, given by $\alpha (t)$ (thick line) and $\beta (t)$ (thin line), close together. ({\bf B}) Distance between bump positions $|\alpha (t) - \beta (t)|$ (dark line) stays small while displacement of bump positions $|\alpha (t)|$ (thick line) and $|\beta (t)|$ (thin line) grows. Threshold $\theta = 0.5$, coupling $w_{12}(x) = w_{21}(x) = \cos (x)$, noise amplitude $\ve = 0.5$, delayed in interlaminar coupling $\tau_{12}(x,y) = \tau_{21}(x,y) \equiv \bar{\tau} = 0.5$.}
\label{sim_alfbet}
\end{figure} 

We now derive effective equations for the positions of the pair of stationary bump solutions in the presence of noise and delayed coupling between layers.  As demonstrated in Fig. \ref{sim_alfbet}, in a typical realization of the dual layer network (\ref{delayers}) with weak noise ($0 < \ve \ll 1$), the distance between bump positions ($|\alpha (t) - \beta (t)|$) remains quite small while their absolute positions $\alpha (t) \approx \beta (t)$ are continually displaced by Brownian motion. We will therefore focus exclusively on the displacement in bump positions, assuming they move together ($\alpha (t) \approx \beta (t) \approx \Delta (t)$). For a detailed analysis that allows different displacements ($\Delta_1$ and $\Delta_2$) for each bump in two weakly coupled layers, see \citep{kilpatrick13c}. Thus, for short enough times ($t \ll 1/ \ve$) we can explore the impact of noise using a perturbation expansion, which assumes the bumps' positions ($\Delta$) and profiles (adding $\Phi_1$ and $\Phi_2$) change, so that
\begin{align}
u_1(x,t) &= U_1(x- \Delta(t)) + \ve \Phi_1( x- \Delta (t),t) + \cdots \nonumber \\
u_2 (x,t) &= U_2(x - \Delta(t)) + \ve \Phi_2 ( x - \Delta (t),t)  + \cdots \label{strexp}
\end{align} 
Such perturbation expansions have been applied to the analysis of stochastic front propagation in nonlinear PDEs \citep{mikhailov83,armero98,garciaojalvo99} and more recently neural field equations \citep{bressloff12b,kilpatrick13}. Substituting the expansion (\ref{strexp}) into (\ref{delayers}), expanding in powers of $\ve$, we find the bump solutions (\ref{bsols}) at ${\mc O}(1)$. At ${\mc O}(\ve)$, we find
\begin{align}
\left( \begin{array}{c} \d \Phi_1 (x,t) \\ \d \Phi_2(x,t) \end{array} \right) - {\mc L} \left( \begin{array}{c} \Phi_1(x,t) \\ \Phi_2(x,t) \end{array} \right) \d t &= \left( \begin{array}{c} \ve^{-1} \d \Delta U_1'(x) + \d W_1(x,t) \\ \ve^{-1} \d \Delta U_2'(x) + \d W_2(x,t) \end{array} \right) + \ve^{-1} {\mc K}(x,t) \label{pexp1}
\end{align}
where ${\mc L}$ is the linear operator
\begin{align*}
{\mc L} \left( \begin{array}{c} u_1(x) \\ u_2(x) \end{array} \right) = \left( \begin{array}{c} - u_1(x) + \int_{- \pi}^{\pi} w(x-y) f'(U_1(y)) u_1(y) \d y +  \int_{- \pi}^{\pi} w_{12}(x-y) f'(U_2(y)) u_2(y) \d y  \\ - u_2(x) + \int_{- \pi}^{\pi} w(x-y) f'(U_2(y)) u_2(y) \d y + \int_{- \pi}^{\pi} w_{21}(x-y) f'(U_1(y)) u_1(y) \d y  \end{array} \right)
\end{align*}
for any vector $\uu (x) = (u_1(x),u_2(x))$ of $L^2$ integrable functions. Reciprocal coupling between the two layers generates the term
\begin{align}
{\mc K}(x,t) = \left( \begin{array}{c} \int_{- \pi}^{\pi} w_{12}(x-y) f'(U_2(y))U_2'(y) ( \Delta(t) - \Delta(t - \tau_{12}(x,y))) \d y \\ \int_{- \pi}^{\pi} w_{21}(x-y) f'(U_1(y))U_1'(y) (\Delta(t) -  \Delta(t - \tau_{21}(x,y)) ) \d y \end{array} \right), 
\end{align}
where the delays are inherited by the stochastic variable representing the bump's position. Note, we have linearized the terms $f(U_j(x+ \Delta (t) - \Delta (t- \tau_{jk}))) = f(U_j(x)) + f'(U_j(x))U_j'(x)(\Delta (t) - \Delta (t - \tau_{jk})) + {\mc O}(|\Delta (t) - \Delta (t- \tau_{jk})|^2)$ ($j=1,2$; $k \neq j$) under the assumption that $|\Delta (t) - \Delta (t- \tau_{jk})|$ remains small. We now enforce a solvability condition for (\ref{pexp1}), requiring that the right hand side is orthogonal to the null space of the adjoint linear operator
\begin{align}
{\mc L}^* \left( \begin{array}{c} p_1(x) \\ p_2(x) \end{array} \right) = \left( \begin{array}{c} - p_1(x) + f'(U_1) \int_{- \pi}^{\pi} w(x-y) p_1(y) \d y + f'(U_1) \int_{-\pi}^{\pi} w_{21}(x-y) p_2 (y) \d y \\ - p_2(x) + f'(U_2) \int_{- \pi}^{\pi} w(x-y) p_2(y) \d y + f'(U_2) \int_{- \pi}^{\pi} w_{12}(x-y)p_1(y) \d y \end{array} \right),  \label{adjop}
\end{align}
for any $L^2$ integrable vector $\p = (p_1(x), p_2(x))^T$ which we have derived using the definition
\begin{align}
\int_{- \pi}^{\pi} \p^T(x) {\mc L}\uu (x) \d x = \int_{- \pi}^{\pi} \uu^T(x) {\mc L}^* \p(x) \d x.  \label{Ladjip}
\end{align}
Identifying the nullspace $(q_1 (x),q_2 (x))$ of ${\mc L}^*$, we can ensure (\ref{pexp1}) is solvable by taking the inner product of both sides of the equation with this vector to yield the equation
\begin{align*}
\langle q_1 , \ve^{-1} \d \Delta U_1' + \d W_1 + \ve^{-1} \int_{- \pi}^{\pi} w_{12}(x-y) f'(U_2(y))U_2'(y) ( \Delta (t) - \Delta (t - \tau_{12}(x,y))) \d y \d t \rangle  + & \\
\langle q_2 , \ve^{-1} \d \Delta U_2' + \d W_2 + \ve^{-1} \int_{- \pi}^{\pi} w_{21}(x-y) f'(U_1(y))U_1'(y) ( \Delta (t) - \Delta (t - \tau_{21}(x,y))) \d y \d t \rangle &= 0,
\end{align*}
defining the $L^2$ inner product $\langle u,v \rangle = \int_{- \pi}^{\pi} u(x) v(x) \d x$ for any $L^2$ integrable functions $u(x)$ and $v(x)$. Therefore, the stochastically evolving bump position $ \Delta (t) $ obeys the delayed stochastic process:
\begin{align}
\d \Delta (t) = \kappa_{12} (\Delta (t- \tau_{12}(x,y))) + \kappa_{21} (\Delta (t- \tau_{21}(x,y))) - ( \bar{\kappa}_{11} + \bar{\kappa}_{22}) \Delta (t)  + \d {\mc W}_1  + \d {\mc W}_2 \label{effsyst}
\end{align}
where coupling results in the terms
\begin{align}
\bar{\kappa}_{jj} = \frac{\left\langle q_j , \int_{- \pi}^{\pi} w_{jk}(x-y) f'(U_k(y))U_k'(y) \d y \right\rangle}{\langle q_1 , U_1' \rangle + \langle q_2 , U_2' \rangle}, \hspace{4mm} j=1,2; \hspace{1mm} k \neq j,  \label{kapjj}
\end{align}
and
\begin{align}
\kappa_{jk} (\Delta(t - \tau_{jk}(x,y))) = \frac{\left\langle q_j , \int_{- \pi}^{\pi} w_{jk}(x-y) f'(U_k(y))U_k'(y) \Delta(t-\tau_{jk}(x,y)) \d y \right\rangle}{\langle q_1 , U_1' \rangle + \langle q_2 , U_2' \rangle}, \hspace{4mm} j=1,2; \hspace{1mm} k \neq j,  \label{kapkj}
\end{align}
and noise impacts the bump positions through the white noise processes $\W (t) = ( {\mc W}_1(t), {\mc W}_2(t))^T$ with
\begin{align*}
{\mc W}_j (t) = \ve \frac{\left\langle q_j(x), W_j(x,t) \right\rangle}{\langle q_1 , U_1' \rangle + \langle q_2 , U_2' \rangle}, \hspace{4mm} j=1,2.
\end{align*}
Note, the white noise terms have zero mean $\langle {\mc W}_j (t) \rangle = 0 $ and diffusive variance so $\langle {\mc W}_j^2 (t) \rangle = D_j t$ ($j=1,2$) with
\begin{align*}
D_j = \ve^2 \frac{\int_{- \pi}^{\pi} \int_{- \pi}^{\pi} q_j(x) q_j(y) C_j(x-y) \d x \d y}{\left[\langle q_1, U_1' \rangle + \langle q_2 , U_2' \rangle\right]^2}, \hspace{4mm} j=1,2,
\end{align*}
and $\langle {\mc W}_1 (t) {\mc W}_2(t) \rangle = D_c t$ with
\begin{align*}
D_c = \ve^2 \frac{\int_{- \pi}^{\pi} \int_{- \pi}^{\pi} q_1(x) q_2(y) C_c(x-y) \d x \d y}{\left[\langle q_1 , U_1' \rangle + \langle q_2 , U_2' \rangle\right]^2}.
\end{align*}

\subsection{Small delay expansion for the effective equations: dual layers}
\label{delou}

We now demonstrate the effectiveness of a small delay expansion in approximating the impact of delays on the stochastic dynamics of bumps, as described by the system (\ref{effsyst}). Note, this was originally developed as a perturbative approximation of a stochastic equations with a single delay \citep{guillouzic99}, but we show this theory applies well to systems of more than one delay \citep{frank05}. To begin, we Taylor expand all functions involving delay, assuming $0 \leq \tau_{jk} \ll 1$, so:
\begin{align}
\kappa_{jk}(\Delta(t- \tau_{jk}(x,y)) ) \d t = \kappa_{jk} (\Delta(t) \d t - \tau_{jk}(x,y) \d \Delta(t) ) + {\mc O}(\tau_{jk}^2), \hspace{4mm} j=1,2; \hspace{1mm} k \neq j,  \label{kapjkexpd}
\end{align}
which means that (\ref{kapkj}) becomes
\begin{align*}
\kappa_{jk} (\Delta(t - \tau_{jk}(x,y)) ) \d t = \bar{\kappa}_{jj} \Delta(t) \d t - {\mc T}_{jk} \d \Delta(t) + {\mc O}(\tau_{jk}^2),  \hspace{4mm} j=1,2; \hspace{1mm} k \neq j,
\end{align*}
where
\begin{align}
{\mc T}_{jk} =  \frac{\left\langle q_j(x), \int_{- \pi}^{\pi} w_{jk}(x-y) f'(U_k(y))U_k'(y) \tau_{jk}(x,y) \d y \right\rangle}{\left\langle q_1 , U_1' \right\rangle + \left\langle q_2 , U_2' \right\rangle}, \hspace{4mm} j=1,2; \hspace{1mm} k \neq j.  \label{taujk}
\end{align}
Keeping only the terms larger than ${\mc O}(\tau_{jk}^2)$, we can approximate (\ref{effsyst}) using the small delay approximation
\begin{align*}
\d \Delta (t) = - {\mc T}_{12} \d \Delta (t) - {\mc T}_{21} \d \Delta (t) + \d {\mc W}_1 + \d {\mc W}_2.
\end{align*}
We can identify how the evolution equation for $\Delta (t)$ has changed by simplifying to find
\begin{align*}
\d \Delta (t) = \frac{\d {\mc W}_1 + \d {\mc W}_2}{1 + {\mc T}_{12} + {\mc T}_{21}},
\end{align*}
so that the mean $\langle \Delta (t) \rangle = 0$ and the variance
\begin{align}
\langle \Delta (t)^2 \rangle = \frac{ D_1 + 2 D_c + D_2}{\left( 1 + {\mc T}_{12} + {\mc T}_{21} \right)^2}t.   \label{effvar}
\end{align}
Thus, we see that the main impact of delays is to reduce the long term variance of bumps' stochastic motion.

\subsection{Calculating nullspace: dual layers}
\label{calcnull}

To compute the effective variance (\ref{effvar}) of bump position, we must find the nullspace of the adjoint operator ${\mc L}^*$ (\ref{adjop}), which satisfies the system
\begin{align*}
 q_1(x) &= f'(U_1) \int_{- \pi}^{\pi} w(x-y) q_1(y) \d y + f'(U_1) \int_{-\pi}^{\pi} w_{21}(x-y) q_2 (y) \d y,  \\
 q_2(x) &= f'(U_2) \int_{- \pi}^{\pi} w(x-y) q_2(y) \d y + f'(U_2) \int_{- \pi}^{\pi} w_{12}(x-y) q_1(y) \d y.
\end{align*}
For a Heaviside firing rate function (\ref{H}), we have that the null vector $(q_1(x),q_2(x))^T$ must satisfy
\begin{align}
 q_1(x) &=  \gamma_a \sum_{x_a=\pm a} \delta (x-x_a) \int_{- \pi}^{\pi} \left[ w(x_a-y) q_1(y) + w_{21}(x_a-y) q_2 (y)\right] \d y, \nonumber  \\
 q_2(x) &= \gamma_b \sum_{x_b = \pm b} \delta (x-x_b) \int_{- \pi}^{\pi} \left[ w(x_b-y) q_2(y) + w_{12}(x_b-y) q_1(y) \right] \d y. \label{Hnulsys}
\end{align}
Therefore, null vector components must be of the form $q_1 (x) = \delta (x + a) + {\mc A} \delta (x -a ) $ and  $q_2 (x) =  {\mc B} \delta (x + b) +  {\mc C} \delta (x -b ) $, where we have divided out the degeneracy guaranteed by rescaling $(q_1,q_2)^T$. Plugging these expressions into the system (\ref{Hnulsys}), we generate the linear system
\begin{align}
 1 &= \gamma_a (w (0) + {\mc A} w (2a) + {\mc B} w_{21} (b-a) + {\mc C} w_{21} (a+b) ) \nonumber \\
 {\mc A} &= \gamma_a (w(2a) + {\mc A} w (0) + {\mc B} w_{21} (a+b) + {\mc C} w_{21} (b-a) ) \nonumber \\
 {\mc B} &= \gamma_b ( w_{12} (b-a) + {\mc A} w_{12} (a+b) + {\mc B} w(0)  + {\mc C} w(2b) ) \nonumber \\
 {\mc C} &= \gamma_b ( w_{12} (a+b) + {\mc A} w_{12} (b-a) + {\mc B} w(2b) + {\mc C} w(0) ).  \label{fourmvnul}
\end{align}
We find the linear system (\ref{fourmvnul}) can be further simplified by taking ${\mc A} = -1$ and ${\mc C} = - {\mc B}$ so that
\begin{align}
1 & = \gamma_a \left[ w(0) - w(2a) + {\mc B} (w_{21} (b-a) - w_{21} (a+b)) \right] \nonumber \\
{\mc B} &= \gamma_b \left[  w_{12} (a+b) - w_{12} (b-a) + {\mc B} (w(0)  - w(2b)) \right].   \label{redBsys}
\end{align}
Now we use the formulas for $\gamma_a$ and $\gamma_b$ given by (\ref{gamab}) to write (\ref{redBsys}) as
\begin{align*}
w_{12} (b-a) - w_{12} (a+b) &= {\mc B}(w_{21} (b-a) - w_{21} (a+b))  \\
{\mc B} ( w_{21} (b-a) - w_{21} (a+b) ) &= w_{12} (b-a) - w_{12} (a+b),
\end{align*}
so we can clearly see that ${\mc B} = [w_{12} (b-a) - w_{12} (a+b)]/[w_{21} (b-a) - w_{21} (a+b)]$. Therefore, the null vector of ${\mc L}^*$ is
\begin{align}
\left( \begin{array}{c} q_1(x) \\ q_2 (x) \end{array} \right) = \left( \begin{array}{c} \delta (x+a) - \delta (x-a) \\ \D \frac{w_{12}(b-a) - w_{12} (a+b)}{w_{21}(b-a) - w_{21} (a+b)} \left( \delta (x+b) - \delta (x-b) \right) \end{array} \right),  \label{q12null}
\end{align}
and note in the symmetric case ($w_{12} \equiv w_{21}$), we have $a \equiv b$ and $q_1(x) \equiv q_2(x) = \delta (x+a) - \delta (x-a)$.

\subsection{Calculating variances: dual layers}
\label{calcvar}

The effective variance (\ref{effvar}) can now be explicitly calculated, assuming a Heaviside firing rate function (\ref{H}) and cosine synaptic weights (\ref{wcos},\ref{intercos}). We can then compare the resulting explicitly computed formulas to the same quantities calculated from numerical simulations. Terms arising due to delay ${\mc T}_{12}$ and ${\mc T}_{21}$ are calculated by first noting the spatial derivative of the bump solutions are $U_1'(x) = - 2( \sin a + M_1 \sin b) \sin x$ and $U_2'(x) = -2 (\sin b + M_2 \sin a) \sin x$. Plugging these formulas along with the null vector (\ref{q12null}) of ${\mc L}^*$ into (\ref{taujk}) and assuming distance-dependent delays $\tau_{jk}(x,y) = \tilde{\tau}_{jk}(x-y) = \tilde{\tau}_{jk}(y-x)$ for $j=1,2$ and $k \neq j$, we find
\begin{align}
{\mc T}_{jk} &= \frac{\cos (b-a) \tilde{\tau}_{jk}(b-a) - \cos (a+b) \tilde{\tau}_{jk} ( a + b) }{2 M_1^{-1} \sin^2 a + 2 \sin a \sin b  + 2 M_2^{-1} \sin^2 b+ 2 \sin a \sin b}.  \label{geneffdel}
\end{align}
Now, to compute the effective diffusion coefficients in each layer, we consider cosine spatial correlations (\ref{coscorr}) and noise that may be correlated ($c_c \geq 0$) between layers. This yields
\begin{align*}
D_1 &= \ve^2 \frac{c_1 \sin^2 a}{4 \left[ \sin^2 a+ M_1 \sin a \sin b  + \sin^2 b + M_2 \sin a \sin b \right]^2}, \\
D_2 &= \ve^2 \frac{c_2 \sin^2 b}{ 4 \left[ \sin^2 a+ M_1 \sin a \sin b  + \sin^2 b + M_2 \sin a \sin b \right]^2}, \\
D_c &= \ve^2 \frac{c_c \sin a \sin b}{4 \left[ \sin^2 a+ M_1 \sin a \sin b  + \sin^2 b + M_2 \sin a \sin b \right]^2}.
\end{align*}

\begin{figure}[tb]
\begin{center} \includegraphics[width=8cm]{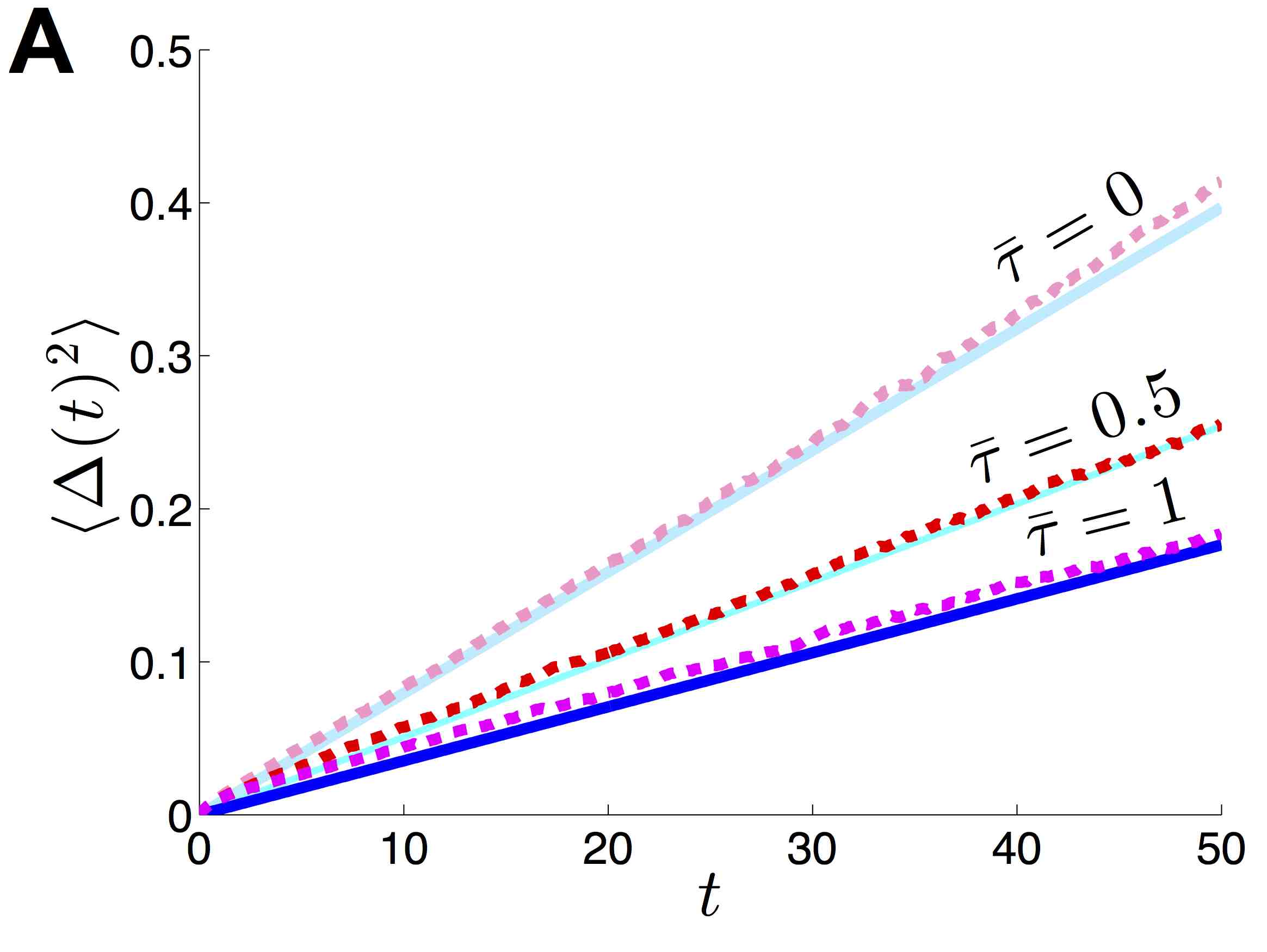}  \includegraphics[width=7.9cm]{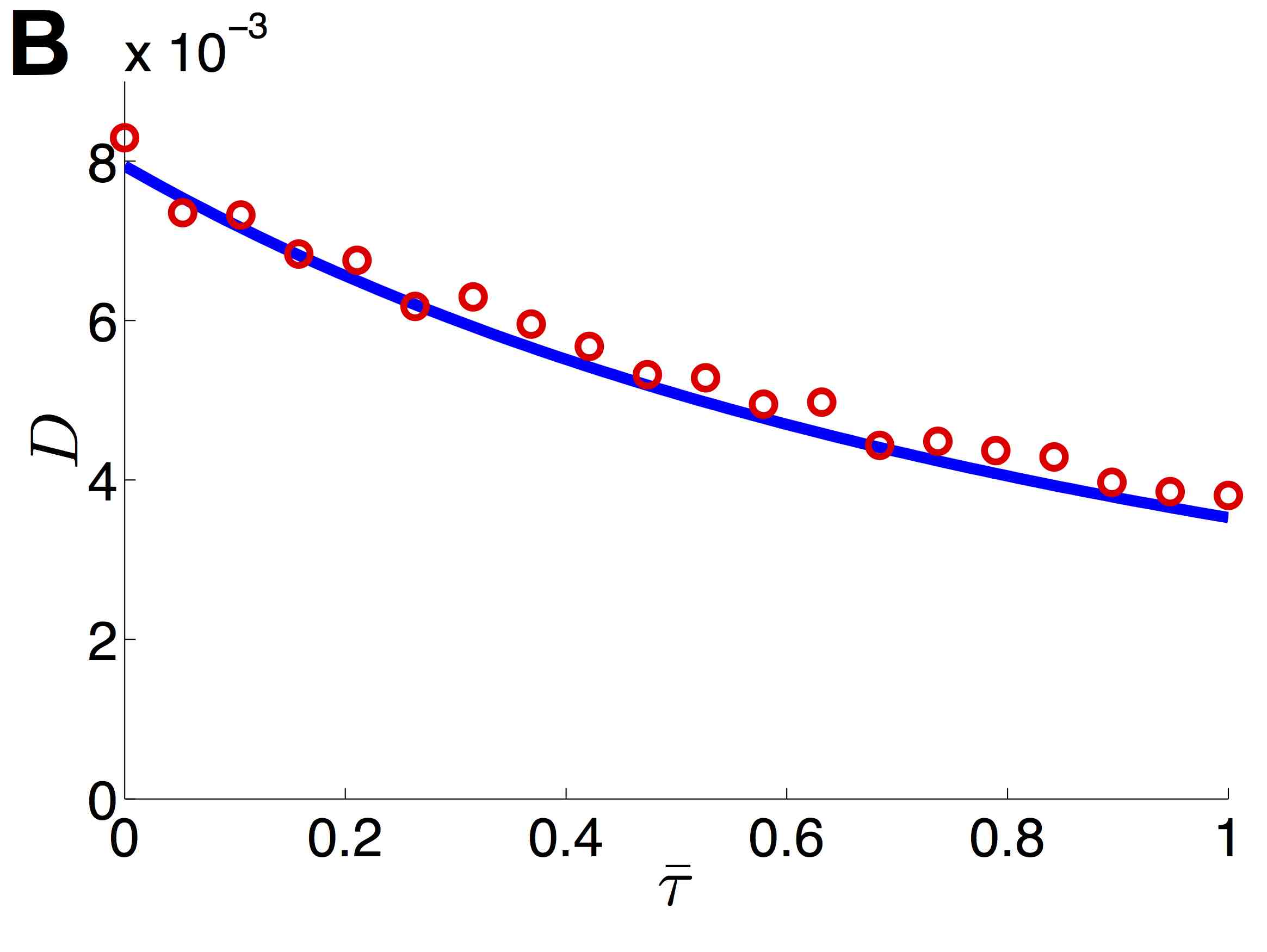} \end{center}
\caption{Effective diffusion $D$ approximated for hard delays $\tau_{12} = \tau_{21} \equiv \bar{\tau}$ and symmetric coupling $w_{12}(x) = w_{21}(x) = \cos (x)$. ({\bf A}) Variance $\langle \Delta (t)^2 \rangle = Dt$ in the position of coupled bumps in a dual layer network coupled with delays (\ref{delayers}) is calculated assuming weak noise and a small delay expansion (\ref{symharddel}). Both our theoretical prediction (solid lines) and numerical simulations (dashed lines) reveal that the effective variance increases more slowly for longer propagation delays $\bar{\tau}$. ({\bf B}) Effective diffusion $D$ decreases as a function of hard delay $\bar{\tau}$ in our asymptotic theory (solid line) and numerical simulations (circles). Threshold $\theta = 0.5$, no noise correlations ($c_c \equiv 0$), noise amplitude $\ve = 0.5$. Variances are computed from numerical simulations using 5000 realizations each.}
\label{nsim_hardsym}
\end{figure}

\begin{figure}[tb]
\begin{center}  \includegraphics[width=8cm]{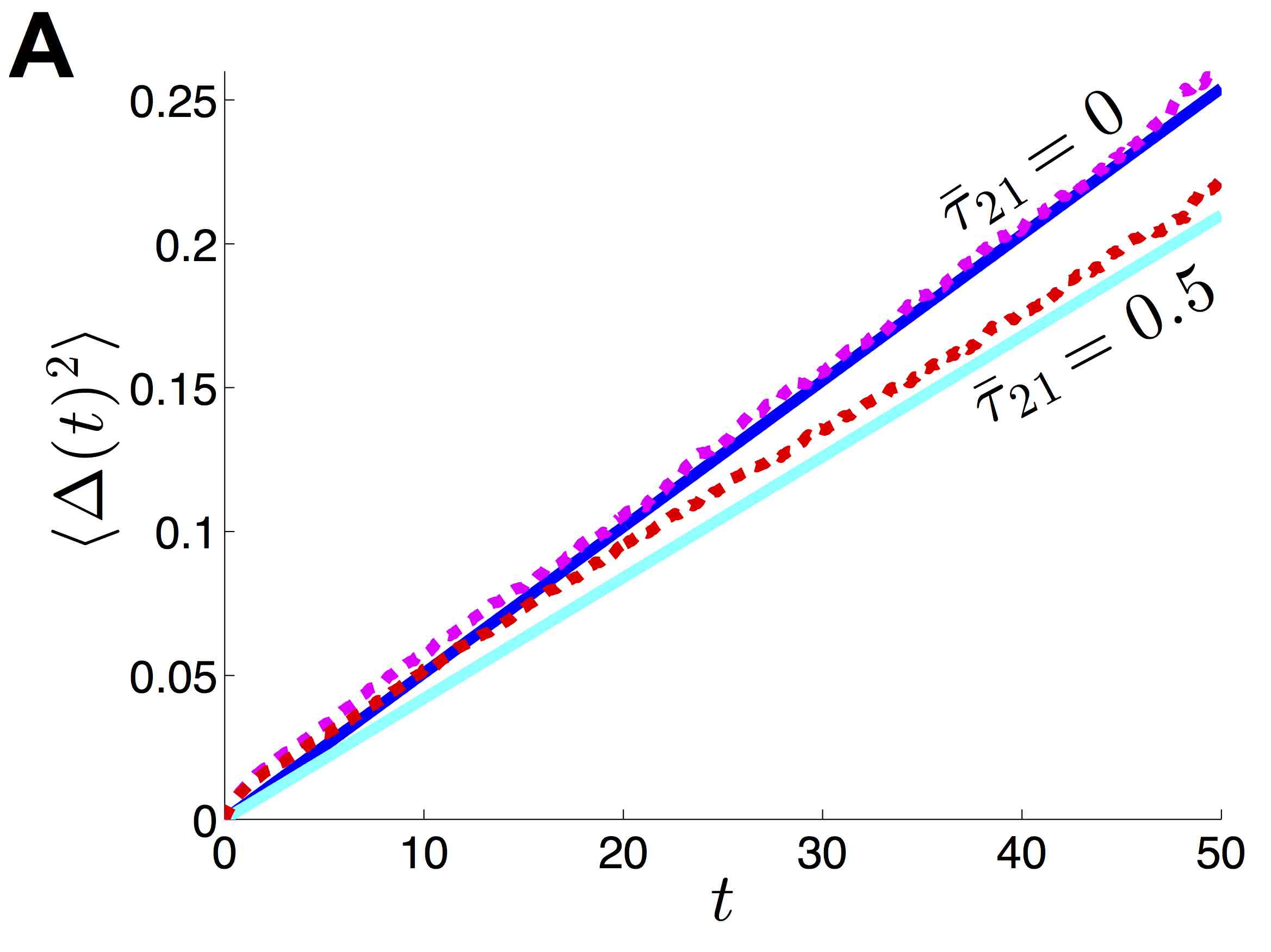}  \includegraphics[width=8cm]{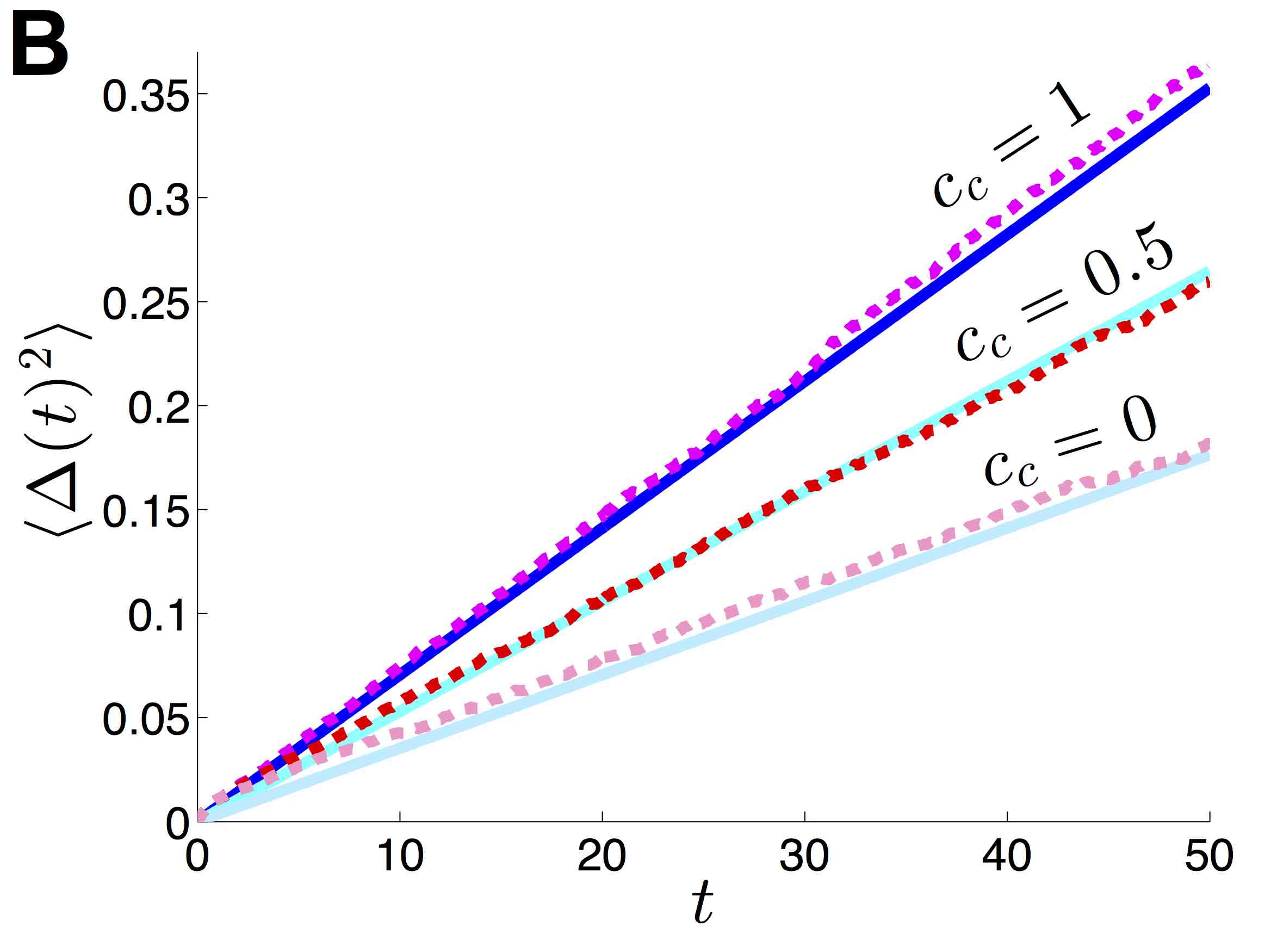} \end{center}
\caption{({\bf A}) The impact of asymmetric hard delays $\bar{\tau}_{21} \neq \bar{\tau}_{12} = 1$ on the variance $\langle \Delta (t)^2 \rangle$ is still well characterized by our theoretical prediction (solid lines) given by (\ref{effvar}) as matched by numerical simulations (dashed lines). ({\bf B}) Our theory (solid lines) predicts variance increases as the amplitude of noise correlations $c_c$ between layers increases (\ref{symharddel}). Threshold $\theta = 0.5$; noise amplitude $\ve = 0.5$; baseline delay $\bar{\tau} = 0$; interlaminar connectivity $w_{12}(x)= w_{21}(x) = \cos (x)$. Variances are computed from 5000 realizations each.}
\label{nsim_corrasym}
\end{figure}

\begin{figure}[tb]
\begin{center}  \includegraphics[width=8cm]{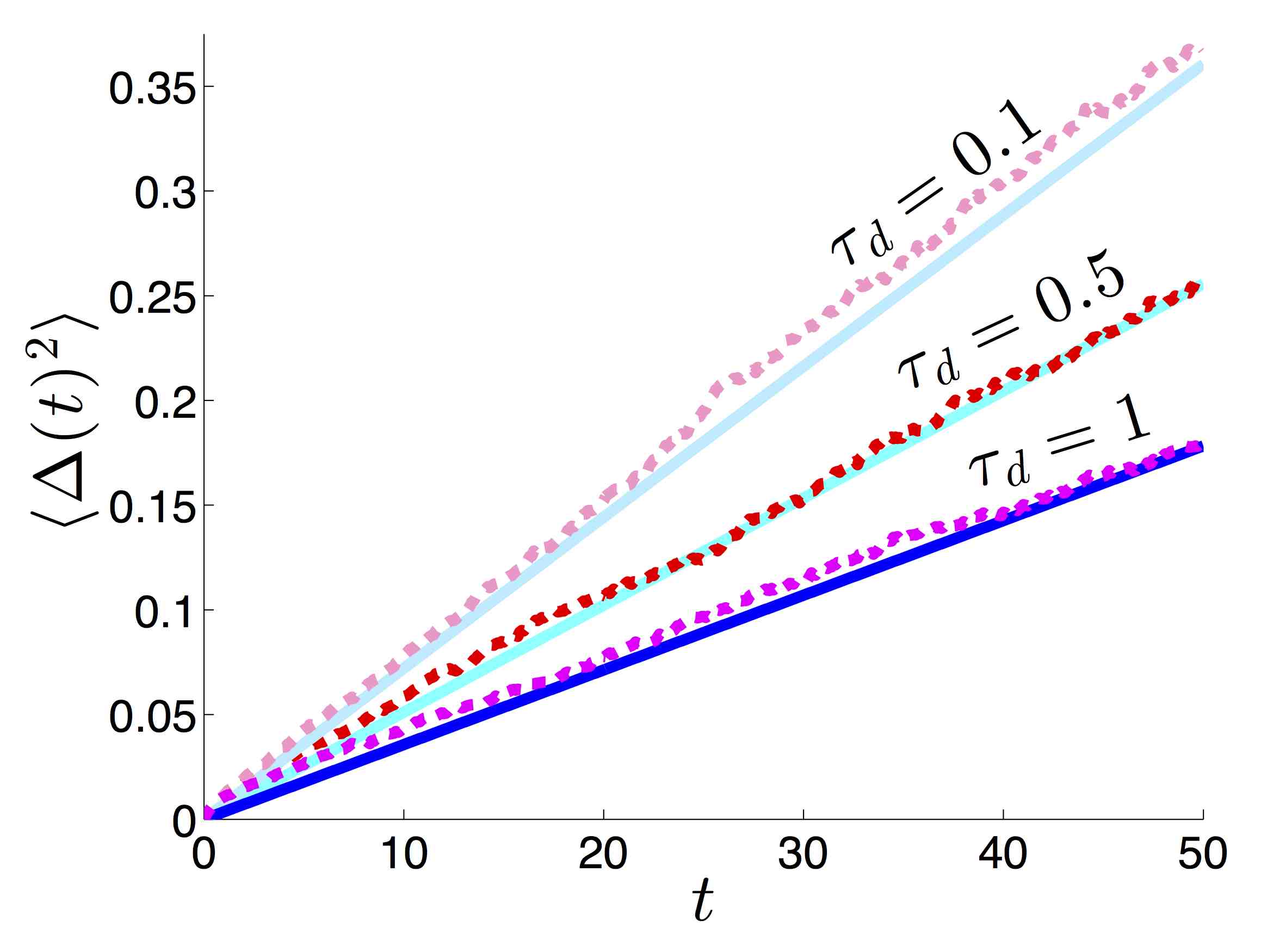} \end{center}
\caption{Distance-dependent delays $\tau_{jk} (x,y) = \tau_d \left[ 1 - \cos (x-y) \right] $ between layers, $(j,k)=(1,2)$ or $(2,1)$, also can stabilize bumps to noise perturbations. Our theoretical calculations (solid lines) suggest that increasing the maximal delay $\tau_d$ further reduces the effective diffusion (\ref{distdepcomp}), which compares well with numerical simulations (dashed lines). Threshold $\theta = 0.5$; noise amplitude $\ve = 0.5$; baseline delay $\bar{\tau} = 0$; interlaminar connectivity $w_{12}(x)= w_{21}(x) = \cos (x)$. Variances are computed from 5000 realizations each.}
\label{nsim_distdep}
\end{figure}

We consider a few different cases of the distance- and layer-dependent delay function $\tilde{\tau}_{jk} (x,y)$. We begin by considering the case where delays are homogeneous in space (hard delays), so $\tau_{jk} =\bar{\tau}_{jk}$, and (\ref{geneffdel}) reduces to
\begin{align*}
{\mc T}_{jk} &= \frac{2 \sin a \sin b \bar{\tau}_{jk}}{M_1^{-1} \left[ 1- \cos (2a) \right] + 2 \sin a \sin b  + M_2^{-1} \left[ 1- \cos (2b) \right] + 2 \sin a \sin b}.
\end{align*} 
To compare our theory to numerical simulations, we begin by focusing on the symmetric case where coupling $M_1 = M_2 \equiv M$, noise $c_1 = c_2 = 1$, and delays $ \bar{\tau}_{12} =  \bar{\tau}_{21} =  \bar{\tau}$, so that $a = b$ and ${\mc T}_{12} = {\mc T}_{21} \equiv {\mc T}$ with
\begin{align*}
{\mc T} = \frac{\bar{\tau}M}{2 (1+M)}.
\end{align*}
In addition, the diffusion coefficients will be identical in each layer $D_1 = D_2 = D_l$ where
\begin{align*}
D_l = \frac{\ve^2}{16 (1+M)^2 \sin^2 a}, \hspace{9mm} D_c = \frac{\ve^2 c_c}{16 (1+M)^2 \sin^2 a}.
\end{align*}
The variance will then be
\begin{align}
\langle \Delta (t)^2 \rangle = Dt = \frac{2 (D_l + D_c)t}{(1+ 2 {\mc T})^2} = \frac{\ve^2 (1+c_c)t}{8 \sin^2 a \left( 1 + M(1+ \bar{\tau}) \right)^2}.  \label{symharddel}
\end{align}
The formula (\ref{symharddel}) demonstrates how the variance is reduced by increases in the delay time $\bar{\tau}$ as well as the coupling strength $M$. We demonstrate the accuracy of this asymptotic approximation in the absence of noise correlations in Fig. \ref{nsim_hardsym}. Furthermore, we show that our asymptotic predictions hold in the case of nonzero noise correlations ($c_c>0$) as well as asymmetric hard delays ($\bar{\tau}_{12} \neq \bar{\tau}_{21}$) in Fig. \ref{nsim_corrasym}. In all cases, longer propagation delays reduce the variance of stochastic bump motion due to their stabilizing effect on bump perturbations.

Next, we consider the impact of distance-dependent delays on the stochastic motion of the coupled bump solution (\ref{bsols}). We model distance-dependence using the periodic function $\tilde{\tau}_{12}(x) = \tilde{\tau}_{21}(x) = \bar{\tau} + \tau_d \left[ 1- \cos (x) \right]$, so when the distance $|x-y|=0$ there is a baseline delay $\bar{\tau}$ and delay increases with distance $|x-y|$. In this case, (\ref{geneffdel}) reduces to
\begin{align*}
{\mc T}_{jk} = \frac{2 (\bar{\tau} + \tau_d (1 - \cos a \cos b) ) \sin a \sin b}{M_1^{-1} \left[ 1- \cos (2a) \right] + 2 \sin a \sin b  + M_2^{-1} \left[ 1- \cos (2b) \right] + 2 \sin a \sin b}, \hspace{9mm} j=1,2; \hspace{3mm} k \neq j.
\end{align*}
Now, for simplicity, we again focus on the symmetric case ($M_1 = M_2 = M$ so $a=b$) to make the effects of distance-dependent delay most transparent in resulting formulas. In this case ${\mc T}_{12} = {\mc T}_{21} = {\mc T}$, and
\begin{align*}
{\mc T} = \frac{M(\bar{\tau} + \tau_d \sin^2 a )}{2(1+ M)},
\end{align*}
so the distance-dependent propagation delay simply adds to the effective hard delay in our asymptotic approximation. The variance is then given by the formula
\begin{align}
\langle \Delta (t)^2 \rangle = Dt = \frac{\ve^2 (1+c_c)t}{8 \sin^2 a \left( 1 + M(1+ \bar{\tau} + \tau_d \sin^2 a) \right)^2}.  \label{distdepcomp}
\end{align}
We demonstrate how the distance-dependent delay reduces the variance in Fig. \ref{nsim_distdep}, matching well with numerical simulations. Thus, we have shown in several examples that propagation delays between layers help stabilize bumps to noise perturbations.

\section{Multiple layered network with delays}
\label{multi}

\subsection{Stationary bumps stabilized by delayed coupling}
We now explore how the principles we have derived for dual layer networks extend to networks with more than two layers. Stationary bump solutions $ (u_1,...,u_N) = (U_1(x),...,U_N(x))$ to the neural field with $N$ layers (\ref{multilayers}) exist in the absence of noise ($\d W_j \equiv 0$, $\forall j$), satisfying the stationary system
\begin{align}
U_j (x) = \int_{- \pi}^{\pi} w(x-y) f(U_j(y)) \d y + \sum_{k =1 , k \neq j}^N \int_{- \pi}^{\pi} w_{jk} (x-y) f(U_k(y)) \d y, \hspace{9mm} j=1,...,N.  \label{multiint}
\end{align}
Again, we fix the threshold crossing points of bumps $U_j (\pm a_j)$ in the case of even symmetric weight functions and a Heaviside firing rate function, converting the implicit integral equation (\ref{multiint}) to an explicit expression for the coupled bump solution
\begin{align*}
U_j(x) = \int_{-a_j}^{a_j} w(x-y) \d y + \sum_{k =1 , k \neq j}^N \int_{-a_k}^{a_k} w_{jk} (x-y) \d y, \hspace{9mm} j=1,...,N.
\end{align*}
To determine the bump half-widths $a_j$, we require self-consistency of $U_j(a_j)= \theta$ to yield the system
\begin{align}
\theta = \int_0^{2a_j} w(x) \d x + \sum_{k=j, k \neq j}^{N} \int_{a_j - a_k}^{a_j + a_k} w_{jk} (x) \d x.  \label{multithresh}
\end{align}
Considering cosine weight functions (\ref{wcos},\ref{intercos}), we can integrate (\ref{multithresh}) to find
\begin{align*}
\theta = 2 \cos (a_j) \left[ \sin (a_j) + \sum_{k=1, k\neq j}^{N} M_{jk} \sin (a_k) \right].
\end{align*}
In the symmetric case $M_{jk} \equiv M$, $\forall j,k$, then $a_j \equiv a$, $\forall a$, and
\begin{align*}
\theta = 2(1+(N-1)M)\cos (a) \sin (a),
\end{align*}
which can be solved to yield a wide ($a_w$) and narrow ($a_n$) bump pair
\begin{align}
a_w = \frac{\pi}{2} - \frac{1}{2} \sin^{-1} \frac{\theta}{1+(N-1)M}, \hspace{9mm} \frac{1}{2} \sin^{-1} \frac{\theta}{1 + (N-1)M}.  \label{multisolbranch}
\end{align}
Thus, increasing the number of layers $N$ will expand the region of parameter space in which one can expect to find bump solutions, as the solution branches (\ref{multisolbranch}) annihilate at $\theta = 1 + (N-1)M$.

Now, we show our stability analysis of bumps in the dual layer network (\ref{delayers}) extends to analysis within a network with an arbitrary number of layers $N$. As before, we employ the expansion
\begin{align}
u_j(x,t) = U_j(x) + \ve \psi_j(x,t) + {\mc O}( \ve^2), \hspace{9mm} \forall j.  \label{multiuexp}
\end{align}
Plugging into (\ref{multilayers}) for $\d W_j \equiv 0$, $\forall j$, and truncating to ${\mc O}( \ve )$, we find
\begin{align}
\dot{\psi}_j(x,t) = - \psi_j(x,t) + \int_{- \pi}^{\pi} w(x-y) f'(U_j(y)) \psi_j(y,t) \d y + \sum_{k \neq j} \int_{- \pi}^{\pi} w_{jk}(x-y) f'(U_k(y)) \psi_k(y, t - \tau_{jk}(x,y)) \d y, \hspace{3mm} \forall j.  \label{multipsi}
\end{align}
Again, neutrally stable solutions are given by the spatial derivative $\psi_j(x,t) = U_j'(x)$, $\forall j$, as can be shown by plugging into (\ref{multipsi}) to yield
\begin{align}
0 &= -U_j'(x) + \int_{- \pi}^{\pi} w(x-y) f'(U_j(y)) U_j'(y) \d y + \sum_{k \neq j} w_{jk}(x-y) f'(U_k(y)) U_k'(y) \d y, \hspace{5mm} \forall j.  \label{Upjeqn}
\end{align}
Differentiating (\ref{multiint}) and integrating by parts, we see that (\ref{Upjeqn}) indeed holds. However, this does not shed light on how delays shape bumps' response to perturbations since there is no explicit timescale attached to the perturbation $\psi_j(x,t) = U_j'(x)$, $\forall j$. To capture the temporal dynamics of the bump solution $u_j = U_j (x)$, $\forall j$, we will examine the evolution of the threshold crossing points for small but arbitrary perturbations.

Assuming a Heaviside firing rate function (\ref{H}), we can compute
\begin{align*}
f'(U_j) = \gamma_j \left[ \delta (x-a_j) + \delta (x+a_j) \right], \hspace{9mm} \forall j,
\end{align*}
where
\begin{align}
\gamma_j^{-1} = |U_j'(a_j)| = |U'(-a_j)| = w(0) - w(2a_j) + \sum_{k \neq j} \left[ w_{jk} (a_k - a_j) - w_{jk} (a_j + a_k) \right].  \label{mgamj}
\end{align}
We then calculate the integrals in (\ref{multipsi}) to find
\begin{align*}
\dot{\psi}_j (x,t) = - \psi_j(x,t) + \gamma_j \sum_{x_j = \pm a_j} w(x-x_j) \psi_j(x_j,t) + \sum_{k \neq j} \gamma_k \sum_{x_k = \pm a_k} w_{jk}(x-x_k) \psi_k(x_j,t - \tau_{jk}(x,x_k)).
\end{align*}
The essential spectrum is associated with solutions satisfying $\psi_j(\pm a_j,t) \equiv 0$, $\forall j, t$, and $\psi_j(x,t) = \e^{-t} \bar{\psi}_j(x)$, $\forall j$, which does not contribute to any instabilities. Other perturbations can be studied by focusing on the evolution of the values $\psi_j( \pm a_j, t)$, satisfying the system of delay differential equations
\begin{align}
\dot{\psi}_j(-a_j,t) &= - \psi_j(-a_j,t) + \gamma_j \sum_{x_j = \pm a_j} w(-a_j - x_j) \psi_j(x_j,t) + \sum_{k \neq j} \gamma_k \sum_{x_k = \pm a_k} w_{jk}(-a_j - x_k) \psi_k( x_k, t - \tau_{jk}(-a_j,x_k)), \nonumber \\
\dot{\psi}_j(a_j,t) &= - \psi_j(a_j,t) + \gamma_j \sum_{x_j = \pm a_j} w(a_j - x_j) \psi_j(x_j,t) + \sum_{k \neq j} \gamma_k \sum_{x_k = \pm a_k} w_{jk} (a_j - x_k) \psi_k (x_k, t - \tau_{jk}(a,x_k)), \hspace{3mm} \forall j. \label{mpsith}
\end{align}
To examine the evolution in bumps' position, in response to perturbations, we can study the evolution of the $2N$ threshold crossing points, given by the equations
\begin{align}
u_j( \pm a_j  + \ve \alpha_j^{\pm} (t), t ) = \theta + {\mc O}( \ve^2), \hspace{5mm} \forall j.  \label{malfthresh}
\end{align}
Taylor expanding (\ref{malfthresh}) and applying (\ref{multiuexp}), we find at ${\mc O}(\ve)$ that
\begin{align}
\alpha_j^{\pm} (t) = \pm \gamma_j \psi_j(\pm a_j, t), \hspace{5mm} \forall j.  \label{alfpsim}
\end{align}
Substituting (\ref{alfpsim}) into (\ref{mpsith}) and focusing on distant-dependent delays $\tau_{jk}(x,y) = \tilde{\tau}_{jk}(|x-y|)$, we find
\begin{align}
\dot{\alpha}_j^-  &= - \alpha_j^- + \gamma_j \left[ w(0) \alpha_j^- - w(2 a_j) \alpha_j^+ + \sum_{k \neq j} \left( w_{jk}(a_k - a_j) \alpha_k^-(t - \tilde{\tau}_{jk}(|a_k - a_j|)) - w_{jk}(a_j + a_k) \alpha_k^+ (t - \tilde{\tau}_{jk}(a_j+a_k)) \right) \right], \nonumber \\
\dot{\alpha}_j^+  &= - \alpha_j^+ + \gamma_j \left[ -w(2a_j) \alpha_j^- + w(0) \alpha_j^+  - \sum_{k \neq j} \left( w_{jk}(a_j + a_k) \alpha_k^-(t - \tilde{\tau}_{jk}(a_j+a_k)) - w_{jk}(a_k - a_j) \alpha_k^+ (t - \tilde{\tau}_{jk}(|a_k - a_j|)) \right) \right], \label{alfjgen}
\end{align}
$\forall j$. As in the case of two layers, we assume the long term bump widths remain the same ($\lim_{t \to \infty} \alpha_j^+ (t) = \lim_{t \to \infty} \alpha_j^-(t)$, $\forall j$). Thus, the long term position of bumps can be identified using the summed variables $\alpha_j(t):= ( \alpha_j^+(t) + \alpha_j^-(t))/2$, $\forall j$. Summing the equations of (\ref{alfjgen}) associated with each $j$, we find
\begin{align}
\dot{\alpha}_j(t) = \sum_{k \neq j} \left[ W_{jk}^- \left( \alpha_k (t - T_{jk}^-) - \alpha_j(t) \right) + W_{jk}^+ \left[ \alpha_k(t - T_{jk}^+ ) - \alpha_j(t) \right) \right],  \hspace{5mm} \forall j, \label{alfjtot}
\end{align}
where $W_{jk}^{\pm} : = \gamma_j w_{jk}(a_k \pm a_j)$ and $T_{jk}^{\pm} := \tilde{\tau}_{jk}(|a_k \pm a_j|)$, $\forall j$. Instantaneous perturbations of the positions $\alpha_j(t)$ will tend to decay slightly when effective delays are positive ($T_{jk}^{\pm} > 0$), so $\lim_{t \to \infty} \alpha_j(t) < \alpha (0)$.

We compute the amount that delays reduce translations of bump position in the case of symmetric coupling ($W_{jk}^{\pm} \equiv W_{\pm}$, $\forall j,k$) and symmetric and hard delays ($T_{jk}^{\pm} \equiv T$, $\forall j,k$). In this case, the system (\ref{alfjtot}) will be a symmetric single lag system
\begin{align*}
\dot{\alpha}_j(t) = W_T \sum_{k \neq j} \left[  \alpha_k(t- T) - \alpha (t) \right], \hspace{5mm} \forall j,
\end{align*}
where $W_T : = W_+ + W_-$. Taking initial conditions $\alpha_j(0) = \alpha_0$ and $\alpha_j(t) = 0$ for $t \in ( - \infty , 0)$, $\forall j$, we can calculate $\alpha_j (T) = \alpha_0 \e^{- (N-1) W_T T}$, $\alpha_j(2T) = \alpha_0 \left( \e^{-3 (N-1)W_T T} + (N-1) W_T T \e^{- 2 (N-1) W_T T} \right)$, iterating to find
\begin{align}
\lim_{t \to \infty} \alpha_j(t) = \alpha_0 (N-1) W_T T \sum_{n=1}^{\infty} \e^{-n (N-1)W_T T} = \frac{\alpha_0 (N-1) W_T T}{\e^{(N-1) W_T T} - 1} < \alpha_0,  \hspace{5mm} \forall j, \label{alfjlimit}
\end{align}
for $(N-1) W_T T>0$. Essentially we find that both increasing the number of layers $N$ as well as increasing the delay time $T$ will decrease the long term impact of a translating perturbation.

\subsection{Effective stochastic motion of bumps in multilayer network}

We can also extend our analysis of the impact of noise on dual layer networks with delayed coupling to the case of the multilayer network (\ref{multilayers}). Our analysis focuses on the stochastic motion of bump position ($\Delta (t)$), and we assume the profiles of the bump in each layer will be perturbed by the noise as well (described by $\Phi_j$, $\forall j$). Thus, we consider the perturbative expansion
\begin{align}
u_j(x,t) = U_j(x - \Delta_j(t)) + \ve \Phi_j(x - \Delta (t), t) + \cdots, \hspace{5mm} \forall j.  \label{multispert}
\end{align}
Substituting (\ref{multispert}) into (\ref{multilayers}), we can expand in powers of $\ve$, finding at ${\mc O}( \ve )$ that
\begin{align}
\d \bphi (x,t) - {\mc L} \bphi (x,t) \d t = \ve^{-1} \U' (x) \d \Delta  + \d \bomega (x,t) + \ve^{-1} {\mc K}(x,t),  \label{multiphie}
\end{align}
where ${\mc L}$ is the linear operator acting on the vector $\bphi (x) = ( \Phi_1(x,t), \cdots , \Phi_N(x,t) )^T$ defined
\begin{align*}
{\mc L} \uu = \left( \begin{array}{c} - u_1 (x) + \int_{- \pi}^{\pi} w(x-y) f'(U_1(y)) u_1(y) \d y + \sum_{k \neq 1} \int_{- \pi}^{\pi} w_{1k}(x-y) f'(U_k(y)) u_k(y) \d y \\ \vdots \\ - u_j (x) + \int_{- \pi}^{\pi} w(x-y) f'(U_j(y)) u_j(y) \d y + \sum_{k \neq j} \int_{- \pi}^{\pi} w_{jk}(x-y) f'(U_k(y)) u_k(y) \d y \\ \vdots \\ - u_N (x) + \int_{- \pi}^{\pi} w(x-y) f'(U_N(y)) u_N(y) \d y + \sum_{k \neq N} \int_{- \pi}^{\pi} w_{Nk}(x-y) f'(U_k(y)) u_k(y) \d y   \end{array} \right),
\end{align*}
for any length $N$, $L^2$-integrable vector of functions $\uu (x) = (u_1 (x) , \cdots , u_N(x))^T $. Spatiotemporal noise is described by the vector $\bomega (x,t) = ( W_1(x,t), \cdots , W_N(x,t))^T$. Delayed coupling between layers is given by the term
\begin{align*}
{\mc K} (x,t) = \left( \begin{array}{c} \sum_{k \neq 1} \int_{- \pi}^{\pi} w_{1k}(x-y) f'(U_k(y)) U_k'(y) ( \Delta (t) - \Delta (t - \tau_{1k}(x,y))) \d y \\ \vdots \\ \sum_{k \neq j} \int_{- \pi}^{\pi} w_{jk}(x-y) f'(U_k(y)) U_k'(y) ( \Delta (t) - \Delta (t - \tau_{jk}(x,y))) \d y \\ \vdots \\  \sum_{k \neq N} \int_{- \pi}^{\pi} w_{Nk}(x-y) f'(U_k(y)) U_k'(y) ( \Delta (t) - \Delta (t - \tau_{Nk}(x,y))) \d y  \end{array} \right),
\end{align*}
delays are inherited by the stochastic variable $\Delta (t)$ for the bump's position. We enforce solvability of (\ref{multiphie}) by requiring the right hand side is orthogonal to the null space of the adjoint linear operator
\begin{align}
{\mc L}^* \pp (x) = \left( \begin{array}{c} - p_1 + f'(U_1) \left[ \int_{- \pi}^{\pi} w(x-y) p_1(y) \d y + \sum_{k \neq 1} \int_{- \pi}^{\pi} w_{k1} (x-y) p_k(y) \d y \right] \\ \vdots \\ - p_j + f'(U_j) \left[ \int_{- \pi}^{\pi} w(x-y) p_j(y) \d y + \sum_{k \neq j} \int_{- \pi}^{\pi} w_{kj} (x-y) p_k(y) \d y \right] \\ \vdots \\ - p_N + f'(U_N) \left[ \int_{- \pi}^{\pi} w(x-y) p_N(y) \d y + \sum_{k \neq N} \int_{- \pi}^{\pi} w_{kN} (x-y) p_k(y) \d y \right] \end{array} \right),  \label{multiadjop}
\end{align} 
for any $L^2$-integrable vector $\pp (x) = ( p_1(x), \cdots , p_N(x))^T$, derived using the inner product definition (\ref{Ladjip}). Upon computing the nullspace $\qq (x) = (q_1(x), \cdots , q_N(x))^T$ of ${\mc L}^*$, we can generate the solvability condition by taking the inner product of both sides of (\ref{multiphie}) with $\qq(x)$ to yield
\begin{align}
\sum_{j=1}^N \langle q_j, \ve^{-1} \d \Delta U_j' + \d W_j + \ve^{-1} \sum_{k \neq j} \int_{- \pi}^{\pi} w_{jk} (x-y) f'(U_k(u)) U_k'(y) ( \Delta (t) - \Delta ( t - \tau_{jk}(x,y))) \d y \d t \rangle = 0, \label{qmultip}
\end{align}
The bump's position will thus evolve according to the delayed stochastic process
\begin{align}
\d \Delta (t) = \sum_{j=1}^N \left[ \sum_{k \neq j} \kappa_{jk}( \Delta (t - \tau_{jk}(x,y))) - \bar{\kappa}_{jj} \Delta (t) + \d {\mc W}_j \right],  \label{delfulmdel}
\end{align}
where coupling between layers generates the terms
\begin{align*}
\bar{\kappa}_{jj} = \frac{\langle q_j , \sum_{k \neq j} \int_{- \pi}^{\pi}w_{jk} (x-y) f'(U_k(y)) U_k'(y) \d y \rangle}{\sum_{j=1}^N \langle q_j, U_j' \rangle}, \hspace{9mm} \forall j,
\end{align*}
and
\begin{align*}
\kappa_{jk} ( \Delta(t - \tau_{jk}(x,y))) = \frac{\langle q_j, \sum_{k \neq j} \int_{- \pi}^{\pi} w_{jk}(x-y) f'(U_k(y)) U_k'(y) \Delta ( t - \tau_{jk}(x,y)) \d y \rangle}{\sum_{j=1}^N \langle q_j, U_j' \rangle }, \hspace{9mm} \forall j,
\end{align*}
and stochasticity arises due to the white noise processes $\W (t) = ( {\mc W}_1(t), \cdots, {\mc W}_N(t))^T$ with
\begin{align*}
{\mc W}_j(t) = \ve \frac{\langle q_j(x), W_j(x,t) \rangle}{\sum_{j=1}^N \langle q_j, U_j' \rangle}, \hspace{9mm} \forall j.
\end{align*}
White noise terms have zero mean $\langle {\mc W}_j(t) \rangle = 0$ and covariance $\langle {\mc W}_j(t) {\mc W}_k(t) \rangle = D_{jk}t$ ($\forall j$, $k \neq j$) with
\begin{align*}
D_{jk} = \ve^2 \frac{ \int_{- \pi}^{\pi} \int_{- \pi}^{\pi} q_j(x) q_k(y) C_j(x-y) \d x \d y}{\left[\sum_{j=1}^N \langle q_j , U_j'  \rangle \right]^2}.
\end{align*}

\subsection{Small delay expansion: multiple layers}

To study the impact of delays on the stochastic motion of bumps, we will employ a Taylor expansion, as in (\ref{kapjkexpd}), that assumes delays are small ($0 \leq \tau_{jk} \ll 1$, $\forall j$, $k \neq j$) so \citep{guillouzic99}
\begin{align*}
\kappa_{jk} ( \Delta ( t - \tau_{jk}(x,y))) \d t = \bar{\kappa}_{jj} \Delta (t) \d t - {\mc T}_{jk} \d \Delta (t) + {\mc O}(\tau_{jk}^2), \hspace{5mm} \forall j, k \neq j,  
\end{align*}
where
\begin{align}
{\mc T}_{jk} = \frac{\langle q_j(x), \int_{- \pi}^{\pi} w_{jk}(x-y) f'(U_k(y)) U_k'(y) \tau_{jk}(x,y) \d y \rangle}{\sum_{j=1}^N \langle q_j , U_j' \rangle}, \hspace{5mm} \forall j, \ k \neq j.  \label{multitaujk}
\end{align}
Keeping only terms larger than ${\mc O}(\tau_{jk}^2)$, we find (\ref{delfulmdel}) becomes
\begin{align*}
\d \Delta (t) = - \left( \sum_{j=1}^N \sum_{k \neq j} {\mc T}_{jk} \right) \d \Delta (t) + \sum_{j=1}^N \d {\mc W}_j. 
\end{align*}
Simplifying, we find
\begin{align*}
\d \Delta (t) = \frac{\sum_{j=1}^N \d {\mc W}_j}{1 + \sum_{j=1}^N \sum_{k \neq j} {\mc T}_{jk}},
\end{align*}
so the mean $\langle \Delta (t) \rangle = 0$ and the variance
\begin{align}
\langle \Delta (t)^2 \rangle = \frac{\sum_{j=1}^N \sum_{k=1}^N D_{jk}}{\left( 1 + \sum_{j=1}^N \sum_{k \neq j} {\mc T}_{jk} \right)^2}t.  \label{multivar}
\end{align}
As before, delays will reduce the long term variance in bumps' stochastic motion, and increasing the number of layers $N$ will further reduce variance.

\subsection{Calculating nullspace: multiple layers}

Now, to compute the variance (\ref{multivar}), we must identify the nullspace of the adjoint operator ${\mc L}^*$ (\ref{multiadjop}), which obeys the system
\begin{align*}
q_1 (x) &= f'(U_1) \left[ \int_{- \pi}^{\pi}w(x-y) q_1(y) \d y + \sum_{k \neq 1} \int_{- \pi}^{\pi} w_{k1} (x-y) q_k(y) \d y \right], \\
& \vdots \\
q_j(x) &= f'(U_j) \left[ \int_{- \pi}^{\pi} w(x-y) q_j(y) \d y + \sum_{k \neq j} \int_{- \pi}^{\pi} w_{kj} (x-y) q_k(y) \d y \right], \\
& \vdots \\
q_N(x) &= f'(U_N) \left[ \int_{- \pi}^{\pi} w(x-y) q_N(y) \d y + \sum_{k \neq N} \int_{- \pi}^{\pi} w_{kN} w(x-y) q_k(y) \d y \right].
\end{align*}
Thus, for a Heaviside firing rate function (\ref{H}), the null vector $\qq (x) = (q_1(x), \cdots , q_N(x))^T$ satisfies
\begin{align}
q_1(x) &= \gamma_1 \sum_{x_1 = \pm a_1} \delta ( x- x_1)  \int_{- \pi}^{\pi} \left[ w(x_1-y) q_1(y) \d y + \sum_{k \neq 1} w_{k1} (x_1-y) q_k(y) \right] \d y, \nonumber \\
& \vdots \nonumber \\
q_j(x) &= \gamma_j  \sum_{x_j = \pm a_j} \delta (x-x_j)  \int_{- \pi}^{\pi} \left[ w(x_j-y) q_j(y) \d y + \sum_{k \neq j} w_{kj} (x_j-y) q_k(y) \right] \d y, \nonumber \\
& \vdots \nonumber \\
q_N(x) &= \gamma_j  \sum_{x_j = \pm a_N} \delta (x-x_N)  \int_{- \pi}^{\pi} \left[ w(x_N-y) q_N(y) \d y + \sum_{k \neq N} w_{kN} (x_N-y) q_k(y) \right] \d y.  \label{mnulqdel}
\end{align}
Null vector components must be of the form
\begin{align}
q_j(x) = {\mc A}_j ( \delta (x + a_j) - \delta (x-a_j) ).  \label{qjsubtract}
\end{align}
Plugging this ansatz into (\ref{mnulqdel}), we can identify an $N \times N$ linear system for the coefficients ${\mc A}_j$ by requiring equality of the coefficients of $\delta (x+a_j)$ (or equivalently $\delta (x-a_j)$) as
\begin{align}
{\mc A}_j &= \gamma_j \left[  {\mc A}_j (w(0) - w(2a_j)) + \sum_{k \neq j} {\mc A}_k (w_{kj}(a_k-a_j) - w_{kj}(a_k+a_j) ) \right].  \label{majlin}
\end{align}
Utilizing the formula for $\gamma_j$ given by (\ref{mgamj}), we can write (\ref{majlin}) as
\begin{align}
{\mc A}_j \sum_{k \neq j} (w_{jk} (a_k-a_j) - w_{jk}(a_k+a_j)) = \sum_{k \neq j} {\mc A}_k (w_{kj}(a_k-a_j) - w_{kj}(a_k+a_j) ).  \label{simpmajlin}
\end{align}
Formulating the linear system in this way, we can see that if interlaminar connectivity is reciprocally symmetric ($w_{jk}(x) = w_{kj}(x)$, $\forall j,k$, then
\begin{align*}
 \sum_{k \neq j} ({\mc A}_j - {\mc A}_k) (w_{jk} (a_k-a_j) - w_{jk}(a_k+a_j)) = 0,
\end{align*}
so that if ${\mc A}_j \equiv 1$, $\forall j$, the linear system is satisfied. More general connection topologies can be addressed by simply breaking the degeneracy of the system (\ref{simpmajlin}) by setting ${\mc A}_1 \equiv 1$ and inverting the resulting $(N-1) \times (N-1)$ linear system. Henceforth, we focus on the symmetric case ($w_{jk} \equiv w_c$, $\forall j,k$), so we have $a_j \equiv a$ and $q_j(x) = \delta (x+a) - \delta (x-a)$, $\forall j$.

\subsection{Calculating variances: multiple layers}

\begin{figure}[tb]
\begin{center} \includegraphics[width=8cm]{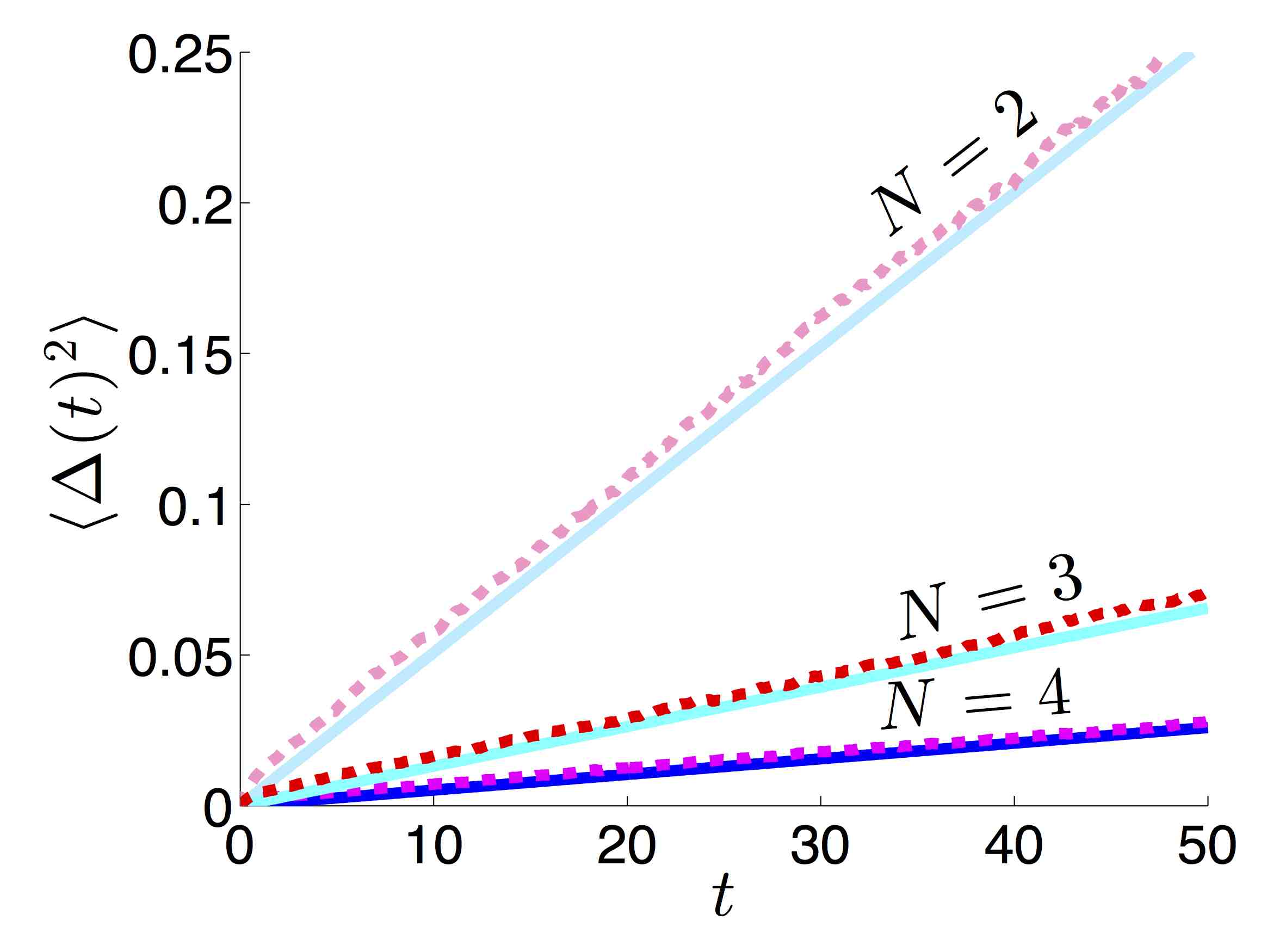} \end{center}
\caption{Effective variance $\langle \Delta (t)^2 \rangle$ in the stochastic motion of bumps in the multilayer stochastic neural field (\ref{multilayers}). We demonstrate how the variance decreases with the number of layers $N$. Our theory (solid lines) reveals that $N$ reduces variance in a divisive way, also scaling the impact of hard delays $\bar{\tau}$ (\ref{multihardvar}), which matches well with numerical simulations (dashed lines). Threshold $\theta = 0.5$; noise amplitude $\ve = 0.5$; delay $\bar{\tau} = 0.5$; interlaminar connectivity $w_{jk} = \cos (x)$, $\forall j$, $k \neq j$. Variances are computed from 5000 realizations.}
\label{multivN}
\end{figure}

We can derive explicit results for the effective variance (\ref{multivar}) by assuming a Heaviside firing rate function (\ref{H}) and cosine synaptic weights (\ref{wcos},\ref{intercos}). We take identical interlaminar connectivity throughout the network ($w_{jk}(x) = M \cos (x)$, $\forall j,k$). Thus, bump half-widths are identical in each layer $a_j \equiv a$, $\forall j$, so $U_j'(x) = -2 (1 + (N-1)M) \sin a \sin x$, $\forall j$. Plugging these expressions along with the null vector (\ref{qjsubtract}) with ${\mc A}_j \equiv 1$, $\forall j$, of ${\mc L}^*$ into (\ref{multitaujk}) and focusing on identical hard delays $\tau_{jk}(x,y) = \bar{\tau}$, $\forall j$, $k \neq j$, we find
\begin{align*}
{\mc T}_{jk} \equiv {\mc T} = \frac{M \bar{\tau}}{N(1+ (N-1)M)}
\end{align*}
Specifying cosine spatial correlations (\ref{coscorr}) and assuming noise to each layer is identical ($c_j \equiv 1$, $\forall j$) and independent ($D_{jk} \equiv 0$, $\forall j$, $k \neq j$), we find that
\begin{align*}
D_{jj} \equiv D_l =  \frac{\ve^2}{4 N^2 (1 + (N-1) M)^2 \sin^2 a}.
\end{align*}
The variance will then be
\begin{align}
\langle \Delta (t)^2 \rangle &= \frac{\ve^2}{4 N \sin^2 a \left[ 1 + (N-1)M(1+ \bar{\tau}) \right]^2}  \label{multihardvar}
\end{align}
As in the case of dual layers, the formula (\ref{multihardvar}) demonstrates that increasing the delay $\bar{\tau}$ will decrease the variance of the bump's stochastic motion. Increasing the number of layers $N$ will decreases the effective variance, as in \citep{kilpatrick13c}. In Fig. \ref{multivN}, we show that our asymptotic prediction of the variance is well matched to the results computed from numerical simulations of the full system (\ref{multilayers}).

\section{Discussion}
\label{disc}

We have shown that propagation delays in the synaptic connections between layers of a neural field can stabilize bumps to noise perturbations. This stabilization utilizes the memory of previous states in other layers provided by delayed coupling. These previous states will be less corrupted by noise, since past states have experienced stochastic forcing for shorter periods of time than the current state. Thus, these past representations of bump position will be a more accurate representation of the initial condition of the network. This provides an additional contribution to the noise reducing mechanism of cancelation, generated by coupling layers together with non-delayed connectivity, as in \citep{kilpatrick13c,kilpatrick14a}. Here, we were able to utilize a small delay expansion to analytically approximate the impact of propagation delays on the effective variance in bump's stochastic motion, showing delays essentially have a divisive effect on variance. We have also extended our previous work by addressing the impact of strong interlaminar coupling upon the stochastic dynamics of bumps, rather than utilizing perturbation theory to explore weak coupling \citep{kilpatrick13c}.

Our work here could be extended in a number of contexts, particularly those concerning the impact of delays on spatial patterns in stochastic neural field equations. First, we plan to explore how propagation delays impact stability of bumps and other patterns in the vicinity of bifurcations. As we have shown here,  lateral inhibitory deterministic neural fields tend to support two co-existent branches of stationary bump solutions, a stable wide bump and an unstable narrow bump, which annihilate in a saddle node bifurcation \citep{amari77,coombes03}. Delays may extend the region in which a stable stationary bump exists in the deterministic system, lengthening the amount of time it would take for noise to generate a rare event whereby the bump is extinguished as in \citep{kilpatrick13}. We will likely need to develop a stochastic amplitude equation approach to study this problem as in \citep{hutt08,kilpatrick14b}. In addition, we plan to explore the impact of delays on propagating patterns, such as traveling waves \citep{kilpatrick14a}. It is questionable whether or not delays will make wave propagation more reliable, since it may lead to instabilities, as in \citep{faye14,laing06}.

\section*{Acknowledgements}

\noindent
This publication was based on work supported in part by the National Science Foundation (DMS-1311755).

\bibliographystyle{elsarticle-num} 
\bibliography{delaybump}

\begin{thebibliography}{10}
\expandafter\ifx\csname url\endcsname\relax
  \def\url#1{\texttt{#1}}\fi
\expandafter\ifx\csname urlprefix\endcsname\relax\def\urlprefix{URL }\fi
\expandafter\ifx\csname href\endcsname\relax
  \def\href#1#2{#2} \def\path#1{#1}\fi

\bibitem{stepan09}
G.~Stepan, Delay effects in brain dynamics, Philosophical Transactions of the
  Royal Society A: Mathematical, Physical and Engineering Sciences 367~(1891)
  (2009) 1059--1062.

\bibitem{stuart97}
G.~Stuart, J.~Schiller, B.~Sakmann, Action potential initiation and propagation
  in rat neocortical pyramidal neurons., The Journal of physiology 505~(Pt 3)
  (1997) 617--632.

\bibitem{vetter01}
P.~Vetter, A.~Roth, M.~H{\"a}usser, Propagation of action potentials in
  dendrites depends on dendritic morphology, Journal of Neurophysiology 85~(2)
  (2001) 926--937.

\bibitem{markram97}
H.~Markram, J.~L{\"u}bke, M.~Frotscher, B.~Sakmann, Regulation of synaptic
  efficacy by coincidence of postsynaptic aps and epsps, Science 275~(5297)
  (1997) 213--215.

\bibitem{izhikevich08}
E.~M. Izhikevich, G.~M. Edelman, Large-scale model of mammalian thalamocortical
  systems, Proceedings of the national academy of sciences 105~(9) (2008)
  3593--3598.

\bibitem{roxin05}
A.~Roxin, N.~Brunel, D.~Hansel, Role of delays in shaping spatiotemporal
  dynamics of neuronal activity in large networks, Physical review letters
  94~(23) (2005) 238103.

\bibitem{bressloff12}
P.~C. Bressloff, Spatiotemporal dynamics of continuum neural fields, J Phys. A:
  Math. Theor. 45~(3) (2012) 033001.

\bibitem{pinto01}
D.~J. Pinto, G.~B. Ermentrout, Spatially structured activity in synaptically
  coupled neuronal networks: I. traveling fronts and pulses, SIAM journal on
  Applied Mathematics 62~(1) (2001) 206--225.

\bibitem{coombes03}
S.~Coombes, G.~J. Lord, M.~R. Owen, Waves and bumps in neuronal networks with
  axo-dendritic synaptic interactions, Physica D: Nonlinear Phenomena 178~(3)
  (2003) 219--241.

\bibitem{hutt03}
A.~Hutt, M.~Bestehorn, T.~Wennekers, Pattern formation in intracortical
  neuronal fields, Network: Computation in Neural Systems 14~(2) (2003)
  351--368.

\bibitem{veltz13}
R.~Veltz, Interplay between synaptic delays and propagation delays in neural
  field equations, SIAM Journal on Applied Dynamical Systems 12~(3) (2013)
  1566--1612.

\bibitem{faye14}
G.~Faye, J.~Touboul, Pulsatile localized dynamics in delayed neural-field
  equations in arbitrary dimension, arXiv preprint arXiv:1402.0530.

\bibitem{laing06}
C.~Laing, S.~Coombes, The importance of different timings of excitatory and
  inhibitory pathways in neural field models, Network: Computation in Neural
  Systems 17~(2) (2006) 151--172.

\bibitem{coombes09}
S.~Coombes, C.~Laing, Delays in activity-based neural networks, Philosophical
  Transactions of the Royal Society A: Mathematical, Physical and Engineering
  Sciences 367~(1891) (2009) 1117--1129.

\bibitem{hutt12}
A.~Hutt, J.~Lefebvre, A.~Longtin, Delay stabilizes stochastic systems near a
  non-oscillatory instability, EPL (Europhysics Letters) 98~(2) (2012) 20004.

\bibitem{abdallah93}
C.~T. Abdallah, P.~Dorato, J.~Benites-Read, R.~Byrne, Delayed positive feedback
  can stabilize oscillatory systems, in: 1993 American Control Conference:
  3106-3107, 1993.

\bibitem{kilpatrick13c}
Z.~P. Kilpatrick, Interareal coupling reduces encoding variability in
  multi-area models of spatial working memory, Frontiers in computational
  neuroscience 7.

\bibitem{funahashi89}
S.~Funahashi, C.~J. Bruce, P.~S. Goldman-Rakic, Mnemonic coding of visual space
  in the monkey's dorsolateral prefrontal cortex, J Neurophysiol. 61~(2) (1989)
  331--49.

\bibitem{wimmer14}
K.~Wimmer, D.~Q. Nykamp, C.~Constantinidis, A.~Compte, Bump attractor dynamics
  in prefrontal cortex explains behavioral precision in spatial working memory,
  Nature neuroscience.

\bibitem{compte00}
A.~Compte, N.~Brunel, P.~S. Goldman-Rakic, X.~J. Wang, Synaptic mechanisms and
  network dynamics underlying spatial working memory in a cortical network
  model, Cereb. Cortex 10~(9) (2000) 910--23.

\bibitem{laing01}
C.~R. Laing, C.~C. Chow, Stationary bumps in networks of spiking neurons,
  Neural Comput. 13~(7) (2001) 1473--94.

\bibitem{amari77}
S.~Amari, Dynamics of pattern formation in lateral-inhibition type neural
  fields, Biol. Cybern. 27~(2) (1977) 77--87.

\bibitem{itskov11}
V.~Itskov, D.~Hansel, M.~Tsodyks, Short-term facilitation may stabilize
  parametric working memory trace, Front. Comput. Neurosci. 5 (2011) 40.

\bibitem{hansel13}
D.~Hansel, G.~Mato, Short-term plasticity explains irregular persistent
  activity in working memory tasks, J Neurosci 33~(1) (2013) 133--49.

\bibitem{camperi98}
M.~Camperi, X.~J. Wang, A model of visuospatial working memory in prefrontal
  cortex: recurrent network and cellular bistability, J Comput. Neurosci. 5~(4)
  (1998) 383--405.

\bibitem{koulakov02}
A.~A. Koulakov, S.~Raghavachari, A.~Kepecs, J.~E. Lisman, Model for a robust
  neural integrator, Nat. Neurosci. 5~(8) (2002) 775--82.

\bibitem{kilpatrick13}
Z.~P. Kilpatrick, B.~Ermentrout, Wandering bumps in stochastic neural fields,
  SIAM J. Appl. Dyn. Syst. 12 (2013) 61--94.

\bibitem{kilpatrick13b}
Z.~P. Kilpatrick, B.~Ermentrout, B.~Doiron, Optimizing working memory with
  spatial heterogeneity of recurrent cortical excitation, submitted.

\bibitem{curtis06}
C.~Curtis, Prefrontal and parietal contributions to spatial working memory,
  Neuroscience 139~(1) (2006) 173--180.

\bibitem{manor91}
Y.~Manor, C.~Koch, I.~Segev, Effect of geometrical irregularities on
  propagation delay in axonal trees, Biophysical Journal 60~(6) (1991)
  1424--1437.

\bibitem{debanne04}
D.~Debanne, Information processing in the axon, Nature Reviews Neuroscience
  5~(4) (2004) 304--316.

\bibitem{bressloff96}
P.~C. Bressloff, New mechanism for neural pattern formation, Physical Review
  Letters 76~(24) (1996) 4644.

\bibitem{wilson73}
H.~R. Wilson, J.~D. Cowan, A mathematical theory of the functional dynamics of
  cortical and thalamic nervous tissue, Biol. Cybern. 13~(2) (1973) 55--80.

\bibitem{goldmanrakic95}
P.~S. Goldman-Rakic, Cellular basis of working memory, Neuron 14~(3) (1995)
  477--85.

\bibitem{ermentrout98}
B.~Ermentrout, Neural networks as spatio-temporal pattern-forming systems,
  Reports on Progress in Physics 61~(4) (1998) 353.

\bibitem{folias11}
S.~E. Folias, G.~B. Ermentrout, New patterns of activity in a pair of
  interacting excitatory-inhibitory neural fields, Phys. Rev. Lett. 107 (2011)
  228103.

\bibitem{bressloff01}
P.~C. Bressloff, Traveling fronts and wave propagation failure in an
  inhomogeneous neural network, Physica D: Nonlinear Phenomena 155~(1) (2001)
  83--100.

\bibitem{coombes04}
S.~Coombes, M.~R. Owen, Evans functions for integral neural field equations
  with heaviside firing rate function, SIAM Journal on Applied Dynamical
  Systems 3~(4) (2004) 574--600.

\bibitem{mikhailov83}
A.~Mikhailov, L.~Schimansky-Geier, W.~Ebeling, Stochastic motion of the
  propagating front in bistable media, Phys. Lett. A 96~(9) (1983) 453 -- 456.

\bibitem{armero98}
J.~Armero, J.~Casademunt, L.~Ramirez-Piscina, J.~M. Sancho, Ballistic and
  diffusive corrections to front propagation in the presence of multiplicative
  noise, Phys. Rev. E 58 (1998) 5494--5500.

\bibitem{garciaojalvo99}
J.~Garc{\'\i}a-Ojalvo, J.~M. Sancho, Noise in spatially extended systems,
  Springer, 1999.

\bibitem{bressloff12b}
P.~C. Bressloff, M.~A. Webber, Front propagation in stochastic neural fields,
  SIAM J Appl Dyn Syst 11~(2) (2012) 708--740.

\bibitem{guillouzic99}
S.~Guillouzic, I.~L'Heureux, A.~Longtin, Small delay approximation of
  stochastic delay differential equations, Physical Review E 59~(4) (1999)
  3970.

\bibitem{frank05}
T.~Frank, Delay fokker-planck equations, perturbation theory, and data analysis
  for nonlinear stochastic systems with time delays, Physical Review E 71~(3)
  (2005) 031106.

\bibitem{kilpatrick14a}
Z.~P. Kilpatrick, Coupling layers regularizes wave propagation in stochastic
  neural fields, Phys Rev E 89~(2) (2014) 022706.

\bibitem{hutt08}
A.~Hutt, A.~Longtin, L.~Schimansky-Geier, Additive noise-induced turing
  transitions in spatial systems with application to neural fields and the
  swift--hohenberg equation, Physica D: Nonlinear Phenomena 237~(6) (2008)
  755--773.

\bibitem{kilpatrick14b}
Z.~P. Kilpatrick, G.~Faye, Pulse bifurcations in stochastic neural fields, SIAM
  J Appl Dyn Syst 13~(2) (2014) 830--860.

\end{thebibliography}

\end{document}